\documentclass[8.5pt,twoside,twocolumn]{article}
\pdfoutput=1
\usepackage[super,sort&compress,comma]{natbib} 
\usepackage{xcolor}
\usepackage{mhchem}
\usepackage{times,mathptmx}
\usepackage[normalem]{ulem}

\usepackage{sectsty}
\usepackage{balance} 

\usepackage{amsmath,graphicx} %eps figures can be used instead
\usepackage{amssymb}
\usepackage{lastpage}
\usepackage[format=plain,justification=raggedright,singlelinecheck=false,font=small,labelfont=bf,labelsep=space]{caption} 
\usepackage{fancyhdr}
\usepackage{siunitx}
\usepackage{graphicx}
\usepackage{etoolbox,hyperref}

\usepackage{verbatim} %ezt adtam hozzá a multiline komment érdekében
\begin{document}

\thispagestyle{plain}
\fancypagestyle{plain}{
%\fancyhead[L]{\includegraphics[height=8pt]{headers/LH}}
%\fancyhead[C]{\hspace{-1cm}\includegraphics[height=20pt]{headers/CH}}
%\fancyhead[R]{\includegraphics[height=10pt]{headers/RH}\vspace{-0.2cm}}
\renewcommand{\headrulewidth}{1pt}}
\renewcommand{\thefootnote}{\fnsymbol{footnote}}
\renewcommand\footnoterule{\vspace*{1pt}% 
\hrule width 3.4in height 0.4pt \vspace*{5pt}} 
\setcounter{secnumdepth}{5}

\newcommand{\comm}[1]{{#1}}
\newcommand{\picomment}[1]{\comm{\color{red} [pi: #1]}}
\newcommand{\ducomment}[1]{\comm{\color{blue} [du: #1]}}

\makeatletter 
\def\subsubsection{\@startsection{subsubsection}{3}{10pt}{-1.25ex plus -1ex minus -.1ex}{0ex plus 0ex}{\normalsize\bf}} 
\def\paragraph{\@startsection{paragraph}{4}{10pt}{-1.25ex plus -1ex minus -.1ex}{0ex plus 0ex}{\normalsize\textit}} 
\renewcommand\@biblabel[1]{#1}            
\renewcommand\@makefntext[1]% 
{\noindent\makebox[0pt][r]{\@thefnmark\,}#1}
\makeatother 
\renewcommand{\figurename}{\small{Fig.}~}
\sectionfont{\large}
\subsectionfont{\normalsize} 

\fancyfoot{}
%\fancyfoot[LO,RE]{\vspace{-7pt}\includegraphics[height=9pt]{headers/LF}}
%\fancyfoot[CO]{\vspace{-7.2pt}\hspace{12.2cm}\includegraphics{headers/RF}}
%\fancyfoot[CE]{\vspace{-7.5pt}\hspace{-13.5cm}\includegraphics{headers/RF}}
%\fancyfoot[RO]{\footnotesize{\sffamily{1--\pageref{LastPage} ~\textbar  \hspace{2pt}\thepage}}}
\fancyfoot[RO]{\footnotesize{\sffamily{\thepage \hspace{2pt} /  \hspace{2pt}\pageref{LastPage}}}}
%\fancyfoot[LO]{\footnotesize{\sffamily{Manuscript submitted to \textit{Acta Materialia}}}}
\fancyfoot[LE]{\footnotesize{\sffamily{\thepage\hspace{2pt} /  \hspace{2pt}\pageref{LastPage}}}}
\fancyfoot[RE]{\footnotesize{\sffamily{Manuscript submitted to \textit{Acta Materialia}}}}
\fancyhead{}
\renewcommand{\headrulewidth}{1pt} 
\renewcommand{\footrulewidth}{1pt}
\setlength{\arrayrulewidth}{1pt}
\setlength{\columnsep}{6.5mm}
\setlength\bibsep{1pt}

\twocolumn[ \begin{@twocolumnfalse}
%\noindent{\textbf{\LARGE Effect of the p$^+$ irradiation induced faults on the acoustic emission, based on in situ Zinc micropillar compression experiments }}

%\noindent{\textbf{\LARGE Microcompression and acoustic emission experiments reveal strain localization and strain burst suppression due to irradiation }}

\noindent{\textbf{\LARGE Irradiation-induced strain localization and strain burst suppression investigated by microcompression and concurrent acoustic emission experiments}}

%\title{Scattering on lipid domains}
\vspace{0.6cm}

{\large \noindent \textbf{Dávid Ugi\textit{$^{a}$}, Gábor Péterffy\textit{$^{a}$}, Sándor Lipcsei\textit{$^{a}$}, Zsolt Fogarassy\textit{$^{b}$}, Edit Szilágyi\textit{$^{c}$}, István Groma\textit{$^{a}$}, and Péter Dusán Ispánovity\textit{$^{a}$}}}\vspace{0.5cm}
%Please note that \ast indicates the corresponding author(s) but no footnote text is required. 

\noindent{\small \textit{$^{a}$~Department of Materials Physics, Eötvös Loránd University, Pázmány Péter Sétány 1/a, H-1117 Budapest, Hungary}}\\
\noindent{\small \textit{$^{b}$~Centre for Energy Research, Institute of Technical Physics and Materials Science, Konkoly Thege M. út 29-33., 1121 Budapest, Hungary}}\\
\noindent{\small \textit{$^{c}$~Institute for Particle and Nuclear Physics, Wigner Research Centre for Physics, Konkoly Thege M. út 29-33., 1121 Budapest, Hungary}}\\


%\noindent\textit{\small{\textbf{Received Xth XXXXXXXXXX 20XX, Accepted Xth XXXXXXXXX 20XX\newline
%First published on the web Xth XXXXXXXXXX 200X}}}

%\noindent \textbf{\small{DOI: 10.1039/b000000x}}
%\vspace{0.6cm}
%Please do not change this text.

\normalsize{ \noindent
Plastic deformation of microsamples is characterised by large intermittent strain bursts caused by dislocation avalanches. Here we investigate how ion irradiation affects this phenomenon during single slip single crystal plasticity. To this end, in situ compression of Zn micropillars oriented for basal slip was carried out in a SEM. The unique experimental setup also allowed the concurrent recording of the acoustic emission (AE) signals emitted from the sample during deformation. It was shown that irradiation introduced a homogeneous distribution of basal dislocation loops that lead to hardening of the sample as well as strain softening due to dislocation channeling at larger strains. With the used deformation protocol strain burst sizes were found to be decreased due to channeling. The concurrently recorded AE events were correlated with the strain bursts and their analysis provided additional information of the details of collective dislocation dynamics. It was found that the rate of AE events decreased significantly upon irradiation, however, other statistical properties did not change. This was attributed to the appearance of a new type of plastic events dominated by short-range dislocation-obstacle interactions that cannot be detected by AE sensors.
}
\vspace{0.5cm}
\end{@twocolumnfalse}]
%Footnotes
%\footnotetext{\dag~Electronic Supplementary Information (ESI) available: [details of any supplementary information available should be included here]. See DOI: 10.1039/b000000x/

%Please use \dag to cite the ESI in the main text of the article.
%If you article does not have ESI please remove the the \dag symbol from the title and the above footnotetext.

%additional addresses can be cited as above using the lower-case letters, c, d, e... If all authors are from the same address, no letter is required

%\footnotetext{\ddag~Additional footnotes to the title and authors can be included \emph{e.g.}\ `Present address:' or `These authors contributed equally to this work' as above using the symbols: \ddag, \textsection, and \P. Please place the appropriate symbol next to the author's name and include a \texttt{\textbackslash footnotetext} entry in the the correct place in the list.}

%ötletem sincs hogy a "]" karakter az előző sorban mire van (twocolumn-hoz tartozik), de anélkül nem tudja lefordítani 

\section{Introduction}

Understanding how irradiation affects the microstructure and mechanical properties of various components is of utmost importance for nuclear applications. Investigations of bulk samples are hindered by the facts that neutron irradiation requires long exposure times and leads to the activation of the sample. For this reason ion irradiation is often used as an alternative since it was shown that the resulting defect structure is comparable to that caused by a neutron flux~\cite{was2002emulation}. However, the limited penetration depth of ions and the resulting inhomogeneity of the defect structure poses a significant challenge for the application of traditional tools of mechanical testing. Recently, therefore, micromechanical approaches (such as nanoindentation or micropillar compression) have been utilised that exploit the fact that ion radiation damage can still be close-to-homogeneous on microscopic scales \cite{Kiener.2011, Albert2020, BUSBY2005267, KATOH2003, landau2014deformation, zhao2015situ, REICHARDT2017323, WEAVER2017368, WANG2018487, cui2021situ, KHIARA2021117096}.

Ion beam changes material structure by radiation damage. In case of FCC crystals these damages mostly appear as stacking-fault tetrahedra (SFTa) with diameters between $1-10{\mathrm{\: nm}}$~\cite{Dai.1997}. These radiation-induced faults obstruct the motion of the active dislocations by acting as pinning centers with a short-range interaction field. This effect was investigated by Kiener et al.~on proton irradiated copper micropillars~\cite{Kiener.2011}. It was found that without any SFTa the deformation took place on multiple and adjacent crystal planes, however, in irradiated samples the deformation became localised into a characteristic single slip band due to shear softening caused by the dissolution of SFTa upon dislocation crossing.

It is well-known that the smallest internal structural element can provide a characteristic length-scale determining the strength of a material~\cite{Arzt.1998}. Consequently, irradiation hardening can be explained based on the Orowan mechanism~\cite{bacon1973r}, so, a dispersed barrier-hardening model~\cite{was2016fundamentals, tan2015formulating} has been introduced to estimate the obstacle-induced changes in yield strength $\Delta\sigma_y$ as:
\begin{equation}
    \Delta\sigma_y=m^{-1}\alpha Gb\sqrt{Nd},
    \label{eqn:hardening}
\end{equation}
where $m$ is the Schmid factor (in case of single crystals), $\alpha$ is a dimensionless strength factor, $G$ is the shear modulus, $b$ is the magnitude of Burgers vector of the moving dislocation, $N$ is the defect density, and $d$ is the typical size of the defects. It is noted that this equation is equivalent to the Taylor hardening low in the sense that here the forest dislocation density is replaced by the dislocation density of loops being equal to $Nd$.

%In the beginning of the XXI century, appeared a possibility of in situ micromechanical testing, which opened a new window for fundamental understanding the mechanical properties of any materials. Since than, numerous paper dealing with the microdeformation mechanisms of metals via microcompression experiments~\cite{Uchic.2004,Uchic.2005,Byer.2010, dimiduk2006scale}. Since the nuclear reactors have become widespread it turned into essential to find fundamental knowledge about irradiation damages. Ion beam changes material structure by radiation damage, but its limited penetration depth requires small-scale testing. In case of fcc crystals the proton beam cases stacking-fault tetrahedra (SFTa) with diameter between $1-10{\mathrm{\: nm}}$~\cite{Dai.1997}. These radiation-induced faults obstruct the motion of the active dislocations like the pinning-depinning dislocation behaviour. It has been investigated by Kiener et al. via proton irradiated copper micropillars~\cite{Kiener.2011}. They found, without any STFs the deformation take place multiple and adjacent crystal planes, however in the irradiated samples the deformation was localised to a characteristic single slip plane. Furthermore, it is well known that the smallest internal structural element can provide a characteristic length-scale determining the strength of a material~\cite{Arzt.1998}. Such a length-scale can form by the average spacing between the irradiation cased stacking faults.

In HCP materials, the anisotropy of the lattice has a strong influence on the evolution of the defect structure. Dislocations with different Burgers vectors are inequivalent in their motion and their reactions with the point defects and their clusters. This causes different type of radiation defects compared to cubic materials. Most results of molecular dynamics simulation work have shown that the interstitial and vacancy clusters form perfect dislocation loops, however their nature (e.g., their plane and Burgers vector) depends on the $\mathrm{c/a}$ ratio \cite{bacon2000primary, tian2021radiation}. In case of zinc basal loops with diameters up to $\mathrm{1\,\mu m}$ are also predicted in the simulations, and observed in TEM investigations as well~\cite{Woo.2000,Mikhin.1997,Whttehead.1978}. Leguey et al.~found that proton irradiation under the fluence of $3 \times 10^{-2}\mathrm{\:dpa}$ in Ti also causes dislocation loops lying in the basal plane and distributed homogeneously in the sample. The fluence only affected the density of these loops~\cite{Leguey.2002,Leguey.2005}. They also reported other dislocations with basal character. These irradiation induced dislocations have a high mobility and only appear in specimens irradiated with a lower fluence.
%These results are in good agreement with earlier investigations reporting that in case of electron irradiated pure zinc with the dose of less than $0.1\mathrm{\:dpa}$ partial dislocation loops also formed in the basal plane~\cite{Griffiths.1993}.

At the micron scale deformation exhibits several characteristic properties different from that of bulk behaviour. As was shown by micropillar experiments conducted on unirradiated single crystalline samples a length-scale similar to the one discussed above is introduced by the specimen size itself. Similarly to Eq.~(\ref{eqn:hardening}) the decrease of this scale (that is, sample size) leads to increased hardness, a phenomenon called size-effect~\cite{Uchic.2004,Uchic.2005,Byer.2010, dimiduk2006scale}. Kiener et al.~argued that for irradiated microsamples the competition of these two length-scales determines the specimen size at which small-scale size-affected plasticity is replaced by bulk behaviour. Consequently, if this transition scale is smaller than the specimen itself then results of micromechanical testing can be directly used to predict bulk properties~\cite{Kiener.2011}.

Another characteristic feature of micron-scale plasticity is that deformation proceeds via intermittent random events (strain bursts). During these events deformation is localized to slip bands and the corresponding deformation increments follow a power-law distribution~\cite{dimiduk2006scale}. These large fluctuations in deformation prevail upon irradiation. Reichard et al.~made in situ micro tensile tests on He$^{2+}$ irradiated pure Ni single crystals~\cite{REICHARDT2015147}. Although the micro sized samples suffered inhomogeneous radiation dose along the cross section, the experiments showed that plastic deformation still occurred via abrupt, non-uniform slip steps that penetrated the full depth of the specimen. These steps were also accompanied by load drops the size of which got larger with increasing irradiation fluence. A similar conclusion was drawn in case of single crystal micropillar compression tests on 304 stainless steel where the plastic deformation was observed to accumulate via the well-known intermittent slips, and next to irradiation-hardening it was also reported that load drops became larger with increasing the irradiation damage~\cite{REICHARDT2017323}. However, discrete dislocation dynamics simulations show that whether irradiation enhance or inhibit strain bursts depends strongly on the level of irradiation as well as on the deformation mode (stress or strain-control) \cite{cui2017does}.

In this paper we aim to employ concurrent acoustic emission (AE) measurements to complement the micromechanical measurements in order to monitor and understand the microscopic processes involved. The AE is generally defined as elastic energy released during local, dynamic and irreversible changes of the microstructure. This released energy forms a stress pulse, which propagates through the material and on the sample's surface and can be detected by a piezoelectric transducer~\cite{Mathis.2012}. Although the industrial usage of AE has become a state-of-the-art method, it is still, in principle, based on empirical results. It is applied to detect cracking, leakage and corrosion with excellent performance~\cite{rettig1976acoustic, dimmick1979acoustical, dunegan1996use}. Nowadays, AE methods are also used for various experiments as a complementary measurement tool that lead to additional and powerful information for research purposes~\cite{Dennett2020, RANGELHERNANDEZ2021, alava2014crackling}. For understanding the connection between dislocation dynamics and AE, James and Carpenter carried out deformation experiments on bulk single crystal zinc and they concluded that the most reasonable physical origin of acoustic signal generation was the sudden dislocation breakaway from pinning points~\cite{james1971relationship}. Since than, several other dislocation processes were identified as source of AE, such as dislocation annihilation~\cite{bolko1980experimental} and twinning~\cite{vinogradov2016limits}. Furthermore, the development of AE technology allowed the in-depth statistical analyses of AE signals. Most importantly, it was shown that the probability distribution of the AE energy $E$ released during individual bursts follows a power law distribution~\cite{miguel2001intermittent, weiss2007evidence} with the scale exponent ${\tau}$ around 1.6 that seems to be independent from the crystal structure~\cite{richeton2006critical}:
\begin{equation}
    P(E) \sim  E^{-\tau}.
\end{equation}
This is in agreement with the distribution of the size of sudden strain bursts during intermittent plastic flow in microsamples~\cite{dimiduk2006scale, maass2015slip, friedman2012statistics} hinting at a close connection between the plastic events and the corresponding AE signals. Indeed, in the recent experiments of Ispánovity et al.~this conjecture was finally proved with the help of a novel micromechanical testing tool equipped with a sensitive AE sensor \cite{ispanovity2022dislocation}. With this set-up AE signals could be detected during the compression of Zn micropillars and, for the first time, a one-by-one match between properties of AE events and those of the corresponding strain bursts could be obtained. The results proved a close-to-linear relationship between the magnitude of the stress drops and the AE energy in an average sense.

%In a more general picture ... criticality, quenched disorder, universality, mild-to-wild fluctuations, etc.

In this paper our aim is to investigate the effects of ion irradiation on the plastic properties of HCP metals. For this purpose single crystalline Zn and p$^+$ implantation is considered and we employ microcompression experiments with concurrent AE measurements that allows us to obtain a more detailed picture of the collective dynamics of dislocations. We mainly focus on how irradiation changes the micromechanical properties (yield stress, strain burst statistics, etc.) and the corresponding AE signals. We also investigate whether irradiation enhances dislocation channeling and corresponding slip band formation as well as its effect on strain burst sizes, and discuss how the AE signals can account for this phenomenon. %The paper is organized as follows... \picomment{TODO}

\section{Sample preparation}

A bulk Zinc single crystal was first mechanically polished using $\mathrm{AlOx}$ particles with an average diameter of $\mathrm{1\:\mu m}$. This was followed by annealing at $\SI{120}{\:\celsius}$ for 5 hours and electropolishing with Struers D2 electrolyte at $10\mathrm{\:V}$ and a maximum current of $2\mathrm{\:A}$ for $\mathrm{10}$ seconds. The surface quality was then verified by electron back-scattered diffraction (EBSD) and the sample was found suitable for proton implantation and the subsequent micropillar fabrication. The top surface was oriented so that its normal formed an angle of $~45^\circ$ with the $\langle 0001 \rangle$ orientation. A perpendicular side surface with a sharp edge was created with FIB milling so that its normal was parallel with the $\langle 2\bar 1 \bar 10 \rangle$ orientation. These surfaces are denoted `top' and `side' in the sketch of Fig.~\ref{fig:srim}a, respectively.

\subsection{Proton implantation}

\begin{figure}[t!]
	\includegraphics[width=0.47\textwidth]{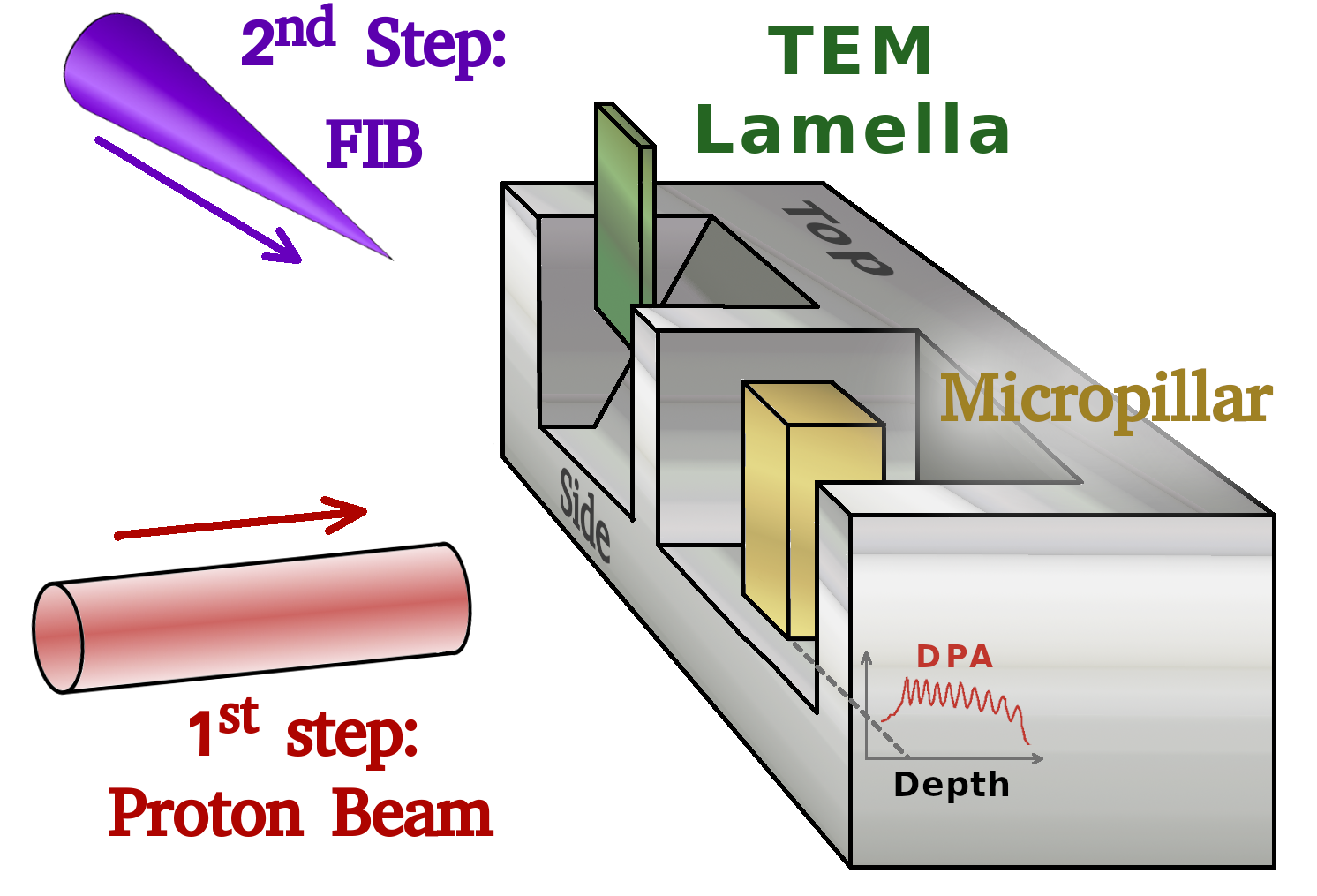}
	\begin{picture}(0,0)
    \put(-250,160){\sffamily{a)}}
    \end{picture}
	\includegraphics[width=0.47\textwidth]{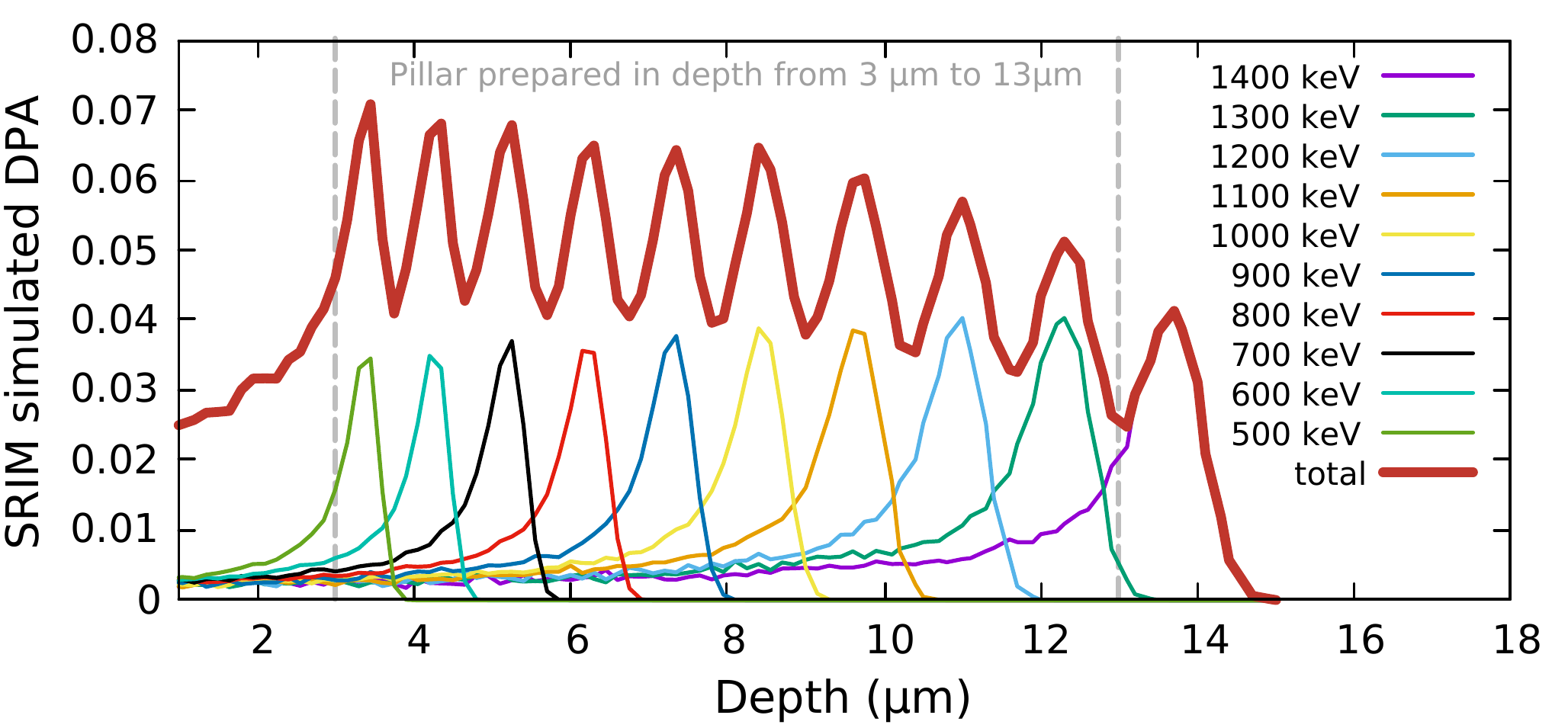}
	\begin{picture}(0,0)
    \put(-250,105){\sffamily{b)}}
    \end{picture}
	\caption{a) Schematic draw of the geometrical arrangement of the experiment. b) DPA profiles calculated with SRIM for the multi-energy p$^+$ irradiation scheme with various fluences summarized in Table \ref{tab:doses}. 
	\label{fig:srim}
	}
\end{figure}
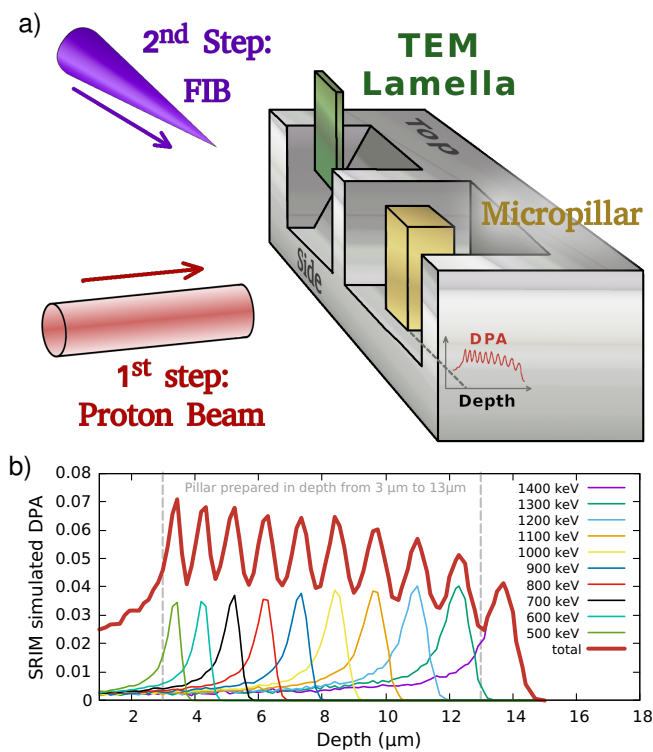

In order to achieve a close-to-homogeneous irradiation damage in a depth of approx.~15 $\mu$m the single crystalline Zn was implanted from the `side' surface (see Fig.~\ref{fig:srim}a) using subsequent p$^+$ beams with energies ranging from 500 keV up to 1400 keV and various fluences. The procedure was performed using the Hungarian Ion-beam Physics Platform accelerator, Institute for Particle and Nuclear Physics, Wigner Research Centre for Physics, Budapest. First, the sample was aligned with respect to the beam. The ion beam of 2 MeV $^4$He$^+$ obtained from a 5 MeV Van de Graaff accelerator was collimated with 2 sets of four-sector slits to the necessary dimensions of $0.5\times0.5$ mm$^2$. To reduce the hydrocarbon deposit during the measurement and the implantation, liquid N$_2$ trap was used. The vacuum in the chamber was approx.~$1.7\times10^{-4}$ Pa. For the calibration, an ion current of typically 10 nA was used as measured by a transmission Faraday cup~\cite{Paszti.1990}. The channel was not found; the random direction for implantation was set to tilt 7$^\circ$ and azimuth 185$^\circ$.

The proton beam was directed on the sample side surface with a beam dimension of $1.5\times1.5$ mm$^2$ such that the center of the spot was approximately at the edge of the sample between the 'top' and 'side' surfaces in Fig.~\ref{fig:srim}a. In order to obtain a more homogeneous beam spot the beam was scanned in both horizontal and vertical directions. The background current of 1-3\% for p was taken into account when determining the necessary dose for a given fluence. The sample was implanted with energies and fluences summarized in Table \ref{tab:doses}. The beam current of typically 80-100 nA was used, except for the two low energy implantation where the current was rather small ($<10$ nA). The error of the dose measurement is around 1\%.

\begin{table}[]
    \centering
    \begin{tabular}{c|c}
        Energy (keV) & Fluence (p/cm$^2$)\\
        \hline
        1400 & $1.6\times10^{16}$ \\
        1300 & $1.5\times10^{16}$ \\
        1200 & $1.4\times10^{16}$ \\
        1100 & $1.3\times10^{16}$ \\
        1000 & $1.2\times10^{16}$ \\
        900 & $1.1\times10^{16}$ \\
        800 & $1.0\times10^{16}$ \\
        700 & $9.5\times10^{15}$ \\
        600 & $8.5\times10^{15}$ \\
        500 & $8.0\times10^{15}$
    \end{tabular}
    \caption{Beam energies and fluences used during the multi-energy implantation.}
    \label{tab:doses}
\end{table}

Damage exposure (i.e., displacements per atom, DPA) profiles were calculated by the SRIM code~\cite{ZIEGLER20101818}, as shown in Fig.~\ref{fig:srim}b. As seen, the fluences for different energies were chosen such that the total DPA profile was close-to-homogeneous in a depth of $3-13$~$\mu$m. Implantations were made at room temperature (RT). Normally, implanting at RT (300 K) causes most of the implantation damage to “self-anneal” since lattice atoms have enough energy to relax into their original crystalline position. However, thermal effects are not considered in SRIM, so the calculated damage corresponds to 0 K implantation. Although thermal effects do change the quantity of final damage, but the same basic damage types are expected to occur.

\subsection{X-Ray measurements}
In order to determine the average dislocation density prior and after irradiation X-ray diffraction (XRD) line profile analysis was used developed by Groma et al.~\cite{Groma.1998, Ungar:gk0172, groma1988asymmetric}. This method is based on the evaluation of the asymptotic behaviour of the $k^{\mathrm{th}}$ order moments of the obtained X-ray diffraction intensity distribution:
\begin{equation}
    v_k(q) = \frac{\int_{-q}^{q} q'^k I(q') dq'}{\int_{-\infty}^{\infty}I(q')dq'}
\end{equation}
where $k$ is the order of the moments, $I(q)$ is the X-ray peak intensity distribution and $q = \frac{2}{\lambda}\left(\sin{\theta}-\sin{\theta_0}\right)$ is the scattering parameter, where $\lambda$ is the wavelength and $\theta$ and $\theta_0$ are the scattering and the Bragg-angle, respectively. In our case the most important moment is the $\mathrm{2}^{\mathrm{nd}}$ order one because it already contains the total dislocation density as a fitting parameter. Its asymptotic form is
\begin{equation}
    v_2(q) = 2 \Lambda \langle \rho \rangle \ln{\left(\frac{q}{\Tilde{q}}\right)}
    \label{eqn:v2}
\end{equation}
where the coefficient $\Lambda$ is related to the dislocation contrast factor and $\Tilde{q}$ is some scaling constant \cite{Groma.1998}. The experimental setup is of $\theta-2\theta$ type, consists of a high intensity Rigaku RU-H3R rotating anode X-ray generator with a copper anode followed by a monochromator that filters out the $\mathrm{K\alpha_2}$ component and redirects the X-ray beam to the sample. The peak profile is recorded with  a Debris MYTHEN2R wide range X-ray detector at a distance of 960 mm. A vacuumed cylindrical chamber between the sample and the detector was also used in order to increase the peak-background ratio.

\begin{figure}[t!]
	\includegraphics[width=0.48\textwidth]{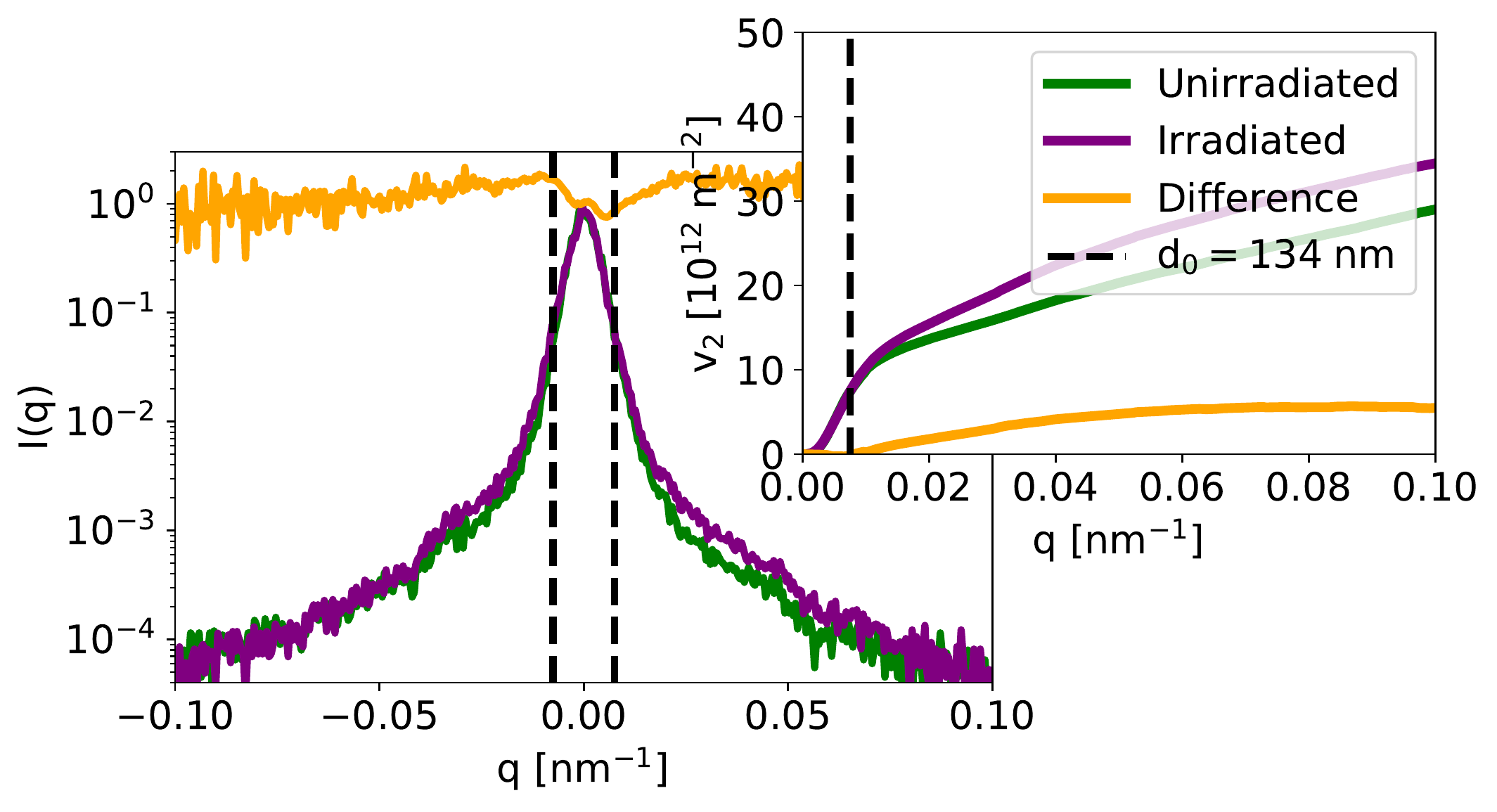}
	\caption{The recorded X-ray intensity peaks corresponding to the diffraction vector $[10\overline{1}0]$. The broadening of the line profiles due to irradiation is clearly observed, as also confirmed by the second order restricted moments shown in the inset. The dashed line corresponds to $q_0 = 1/d_0 = (135 \; \mathrm{nm})^{-1}$ where the restricted moments start to separate. The inset shows the 2nd order restricted moments of the two profiles. The difference of the profiles and the restricted moments is also plotted with orange colour.
	\label{fig:xraypeak2}}
\end{figure}

The line profiles obtained before and after irradiation as well as the corresponding 2nd order restricted moments are shown in Fig.~\ref{fig:xraypeak2}. From fitting Eq.~(\ref{eqn:v2}) one obtains that the total dislocation density increased from $\langle \rho_\mathrm{init}\rangle = 1.1 \times 10^{14} \; \mathrm{m^{-2}}$ to $\langle \rho_\mathrm{irrad}\rangle = 1.8 \times 10^{14} \; \mathrm{m^{-2}}$ due to the irradiation, which clearly indicates the appearance of a significant amount of new dislocation loops. Nevertheless, one can also obtain information about the size of the loops, by observing the difference between the second order restricted moments. It is clear, that around $q_0 = \mathrm{0.0075 \; nm^{-1}}$, which is equivalent to $d_0 = 1/q_0 = \mathrm{135 \; nm}$, the moments of the irradiated and unirradiated sample begin to separate, meaning that the effect of the appearance of the loops on the profile is negligible below the beforementioned limit. This suggests, that the characteristic size of the radiaton-induced loops is expected to be less or equal to $d_0 = \mathrm{135 \; nm}$.

\subsection{Micropillar fabrication}

Micropillars with a $10 \times 10 \mathrm{\:\mu m}^2$ square cross-section were fabricated with a height to side aspect ratio of 3:1. It was shown earlier that this size is already sufficient to allow the detection of a large number of AE signals \cite{ispanovity2022dislocation}. The microsamples were prepared using a focused ion beam (FIB) in a FEI Quanta 3D dual-beam scanning electron microscope (SEM) with a methodology that allows precise control over the resulting sample geometry~\cite{Hegyi.2017, ispanovity2022dislocation}. During milling the ion current is reduced after each step to optimise the duration of the process and minimize the Ga$^+$ ion contamination. The final ion current was set to 500 pA that resulted in rectangle shape pillars with opposite faces being sufficiently parallel, so its cross-section area was identical along the height of the micropillar.

It is mentioned that EBSD measurements showed the unexpected appearance of a small number of twin crystals caused by irradiation. These twins had lamella like shape with a maximum of $10 \mathrm{\:\mu m}$ thickness, and a lateral extension of $100\mathrm{\:\mu m}$. Spacing between these lamellae were a couple of $0.1$~mm, and the micropillars were fabricated from these parts which preserved the original crystal orientation.

%\picomment{Itt még le kellene írni, hogy pontosan hogyan is lettek faragva az élre a pillárok, honnan jött a sugárzás meg ilyesmi. Ehhez kellene még egy szép 3D ábra is, ami a Fig.~1-hez passzolna.}

\begin{figure}[t!]
\begin{center}
	\includegraphics[width=0.4\textwidth]{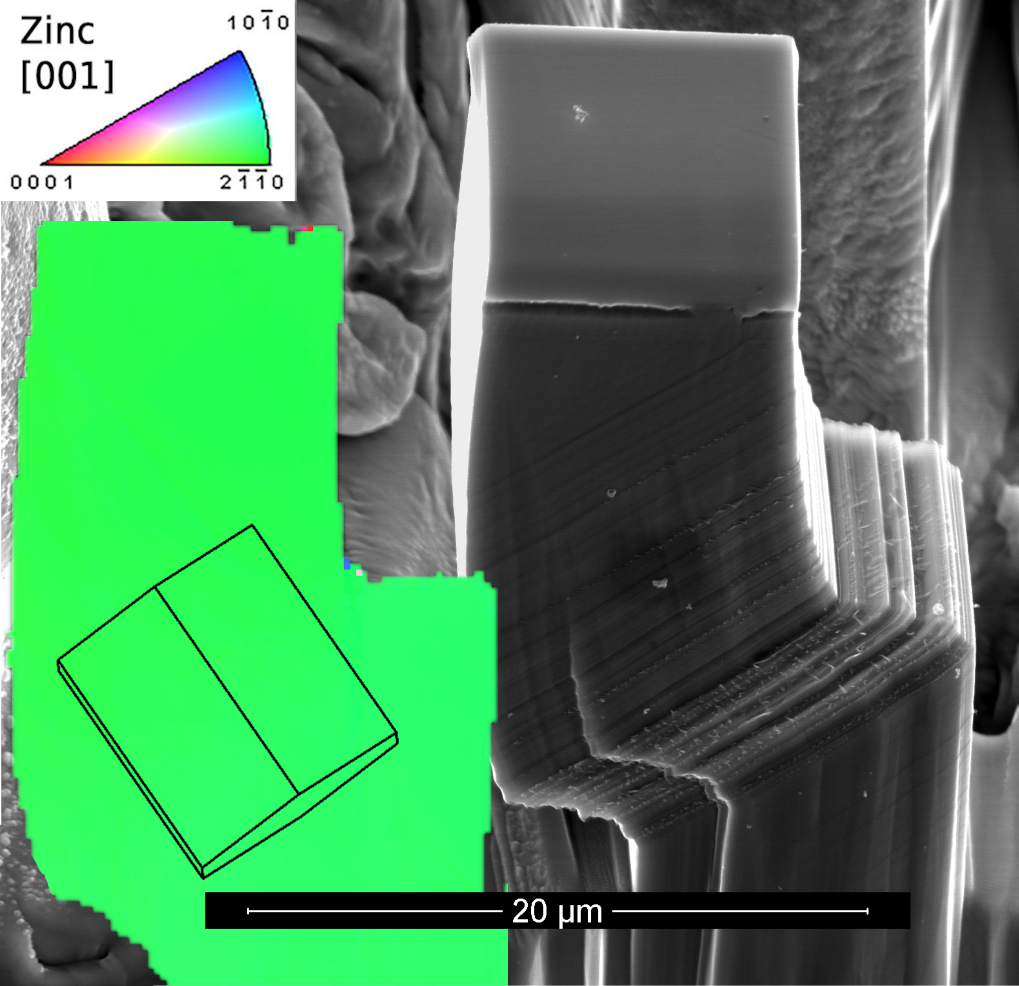}
	\caption{A representative SEM image of one of the unirradiated micropillars taken after compression. The embedded coloured figure on the left shows its EBSD map that confirms that the deformation took place by purely dislocation glide on the basal plane. Note, that slight FIB polishing was performed on the sample surface prior to EBSD measurements that removed $\sim10$ nm in order to eliminate the irregularities of the surface caused by the slip bands. This procedure was performed in order to obtain better quality EBSD maps and did not affect the crystal orientation that the measurement intended to identify.}
	\label{fig:ebsd}
\end{center}
\end{figure}

In total 8 micropillars were fabricated with identical geometry, 4 in the unirradiated region of the sample and 4 in the irradiated region. All of them, therefore, had the same crystal orientation, with the basal plane having an approximately 45$^\circ$ inclination with respect to the loading axis to favor dislocation glide on the basal plane. The pillars were fabricated on the edge of the sample (Fig.~\ref{fig:srim}a) which allows us to obtain EBSD maps directly from the surfaces of the pillars. In addition, this geometry resulted in the close-to-homogeneous irradiation damage throughout the volume of the pillars in the irradiated case. The crystal orientation was unchanged during the compression as shown by the EBSD map taken after the compression in Fig.~\ref{fig:ebsd} which confirms that twinning did not take place during deformation.

%As known the properties of the AE in this unirradiated sample by previous measurements \textcolor{cyan}{cite avalance}, the diameter of the pillars were chosen of $10\mathrm{\:\mu m}$, and the aspect ratio was the commonly applied 3:1 of height to side. For the focused ion beam (FIB) milling method, a FEI Quanta 3D dual beam scanning electron microscope (SEM/FIB) was used in the laboratory of ELTE. For the micro pillar preparation procedure, we chosen the method that developed by Hegyi et al.~\cite{Hegyi.2017}. (More details in \textcolor{cyan}{cite avalance}) In this method we reduce the ion current step by step to optimise the duration of the process and minimize the $Ga^+$ ion implantation. The final ion current was set to $500\mathrm{\:pA}$ what resulted a rectangle shape pillars that the opposite faces are sufficiently parallel, so it's cross-section area independent from the depth. 

\subsection{Transmission electron microscopy}

\begin{figure}[t!]
\begin{center}
	\includegraphics[width=0.35\textwidth]{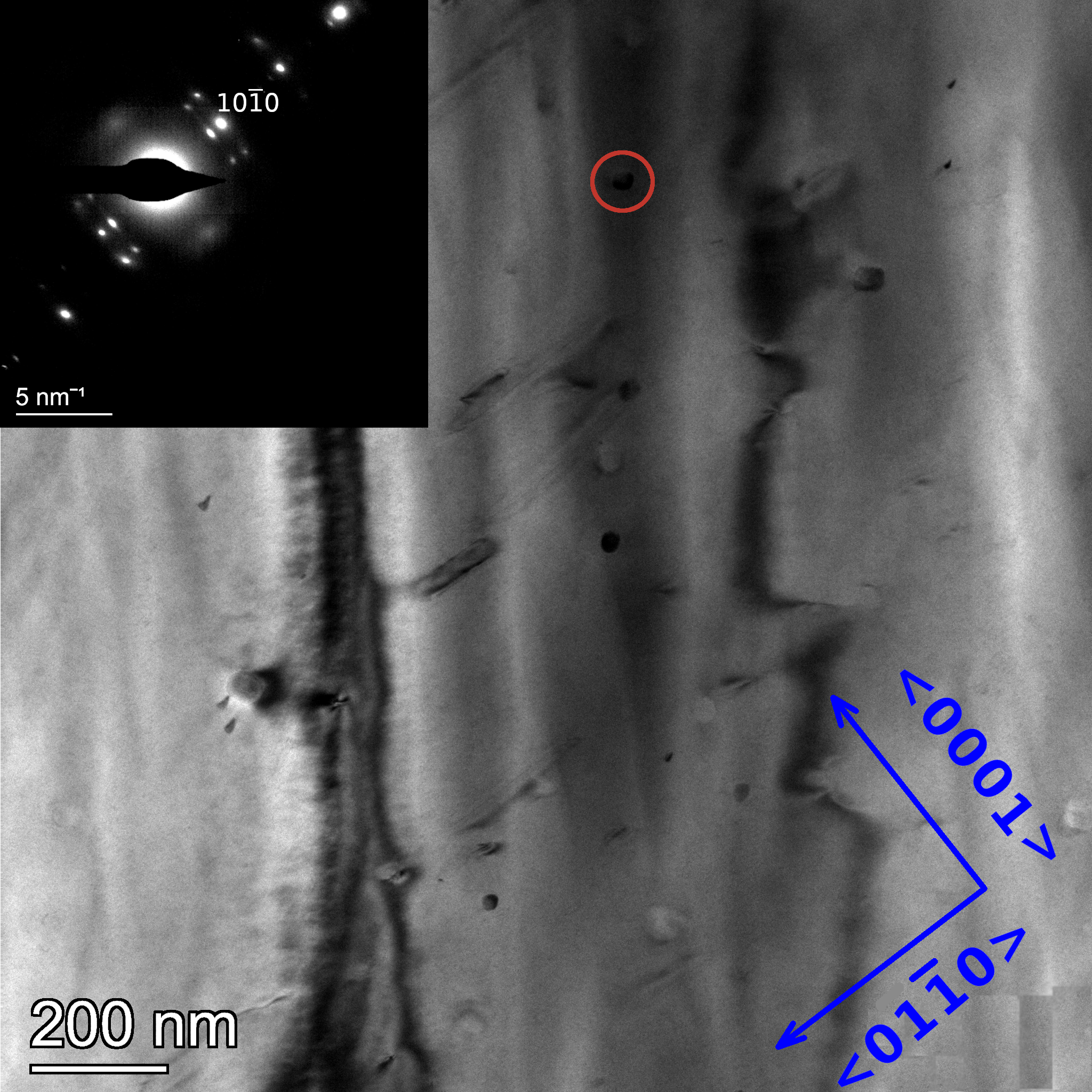}
	\begin{picture}(0,0)
    \put(-195,170){\sffamily{a)}}
    \end{picture}
	\includegraphics[width=0.35\textwidth]{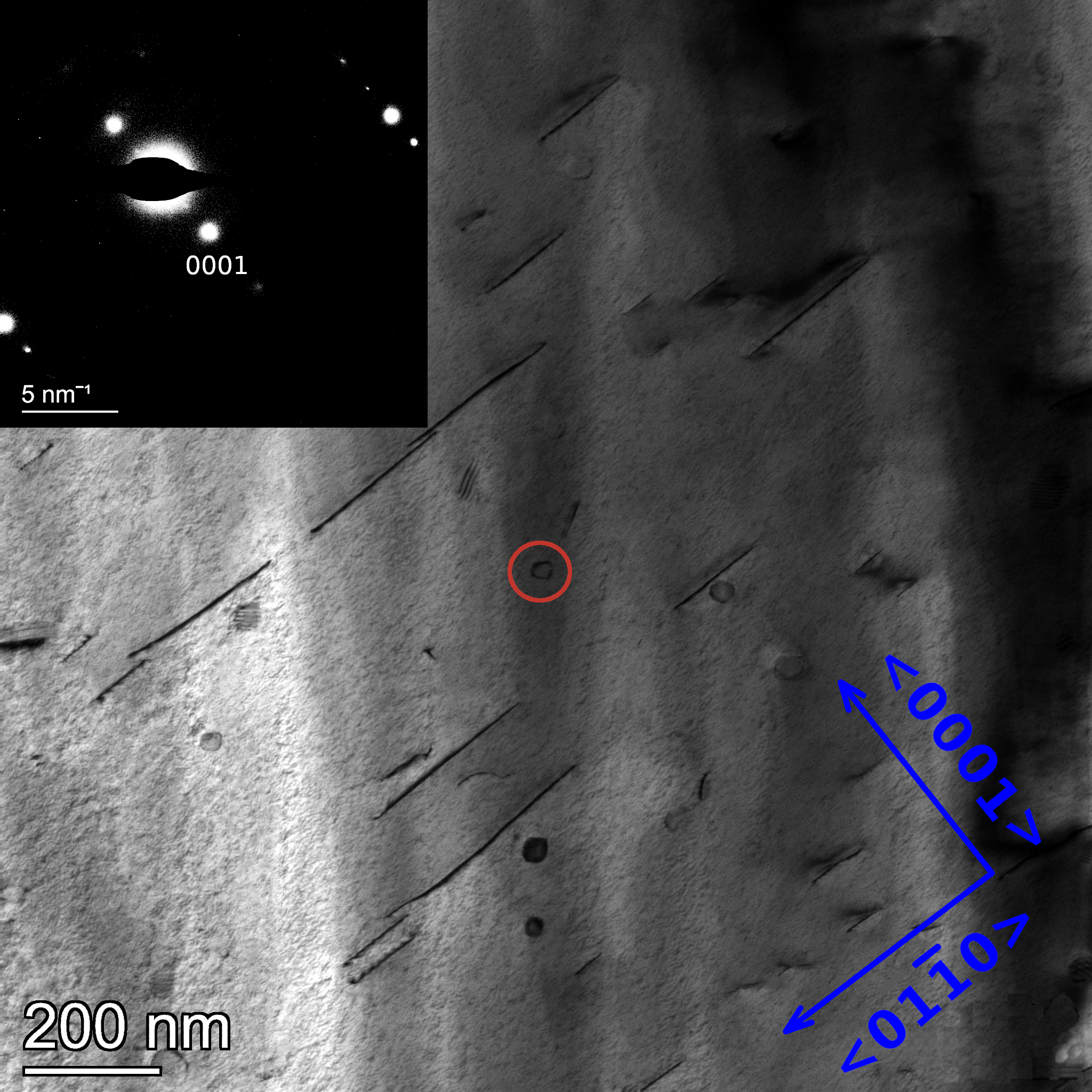}
	\begin{picture}(0,0)
    \put(-195,170){\sffamily{b)}}
    \end{picture}
	\caption{TEM images in dark field imaging mode from an irradiated area showing the cross sections of irradiation-induced dislocation loops. The two images were made with different tilting angles in order to characterize the dislocation types (see text). The two images are slightly shifted, and the red circles mark the same spots of the sample.}
	\label{fig:TEM}
\end{center}
\end{figure}
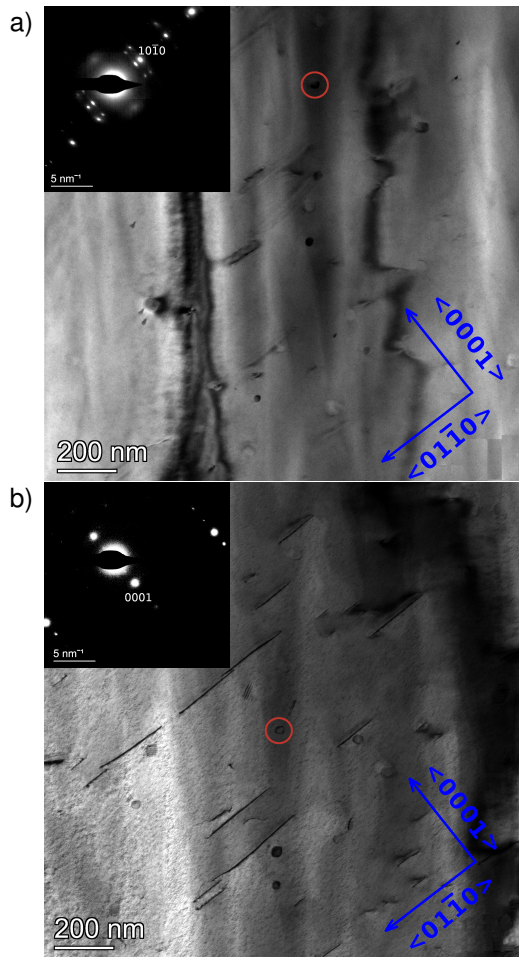

The TEM investigation was carried out using a Cs-corrected TEM Themis-type electron microscope with an operation voltage of 200~keV. The TEM lamella was prepared by FIB thinning from the sample's top surface such that its normal direction was parallel to the front side of the micropillars as seen in the sketch of Fig.~\ref{fig:srim}a. The thickness of the lamella was $45\pm15\mathrm{\:nm}$. The investigated area (Fig.~\ref{fig:TEM}) was in the depth of $\sim$10$\mathrm{\:\mu m}$ where the DPA was the same as in the pillars. Local microstructure was investigated by tilting the specimen first to the $(10\overline{1}0)$ (Fig.~\ref{fig:TEM}a) and then to the $(0001)$ (Fig.~\ref{fig:TEM}b) two-beam positions. In case of the $(0001)$ two-beam condition, the dislocation loops appear as dark lines on the basal plane, that remained in fix positions during the investigation. Since, these dark lines on the basal plane disappear in Fig.~\ref{fig:TEM}a, it means that the Burgers vector of these fix dislocations are of $\langle 0001 \rangle$ type. Such loops may form due to the aggregation of vacancies or interstitials on the basal plane causing a stacking fault and a Schockley-type partial dislocation at its edge. The basal plane is preferred due to the relatively low stacking fault energy due to the large c/a ratio in Zn. The growth of such loops is limited by the energy contribution of the stacking fault that is proportional with the area of the loop. The reason for the existence of these relatively large loops may be the formation of double layer vacancy or interstitial clusters as was observed for irradiated Mg \cite{xu2017origin}. We note that it has been seen during other, so far unpublished, experiments that mobile edge dislocations lying on the basal plane can easily come into motion due to the heating effect of the electron beam which explains the absence of the pre-existing dislocations in the TEM images. %\textcolor{cyan}{\href{https://metalog.elte.hu/s/AXYtQwCSqxEwjEx}{CLICK!}.}} . 

%\picomment{Itt nem tudunk betenni egy olyan képet, ami a hosszú eredeti bazális diszlokációkat mutatja?}
%As we discussed previously, the originally existed mobile edge dislocations were disappeared from the investigated area in the duration about a minute, if them line, parallel to the electron beam of the TEM (The same arrangement as ours). To assume, after this short incubation time the presented dislocation were the result of the irradiation, we can calculate the increase of the dislocation density by irradiation. 

To quantify the increase in the dislocation density due to irradiation the TEM images showing the radiation induced loops were investigated. Based on Fig.~\ref{fig:TEM}a $\sim$36 dislocation loops can be found in the basal plane with an average diameter of $105\pm87\mathrm{\:nm}$ (in agreement with the characteristic maximum diameter of $d_0 = 135 \; \mathrm{nm}$ suggested by X-ray diffraction), and 12 additional circular smaller loops with an average diameter of $37\pm9\mathrm{\:nm}$ that are speculated to lie either on the prismatic or pyramidal plane. Since the volume investigated by TEM is approx.~$0.15\pm0.05\mathrm{\:\mu m^3}$, we find that the dislocation density corresponding to the visible dislocation loops is $\langle \rho_\mathrm{loop} \rangle = 8.3\times10^{13}\mathrm{\:m^{-2}}$. This value is in an agreement with the results of the X-ray measurements discussed above where a difference in the dislocation density between the irradiated and the unirradiated regions was found to be $\langle \rho_\mathrm{irrad} \rangle - \langle \rho_\mathrm{init} \rangle \approx 7\times10^{13}\mathrm{\:m^{-2}}$. It must be noted that the cross section of the X-ray beam was almost two times larger than the irradiated area that can explain the slight difference between the two measured values.

\section{In situ experimental methods}

\subsection{Micromechanical experiments}
\label{sec:micromechanical_experiments}

%It was monitoring the sample during the steps of preparation by EBSD. It had noticed, that twin crystalline formed in single crystal caused by irradiation. These twins have lamella like shape at maximum $10\mathrm\:\mu m$ thickness, and more than $100\mathrm\:\mu m$ length. Spacing these lamella are a couple of $0.1\mathrm\:mm$, so the irradiated part of the sample also suitable for micropillar fabrication in the original crystal orientation. However these twins formation were unexpected in this irradiation dose based on \textcolor{cyan}{Section 1.3} it is possible that the STFs were served as a source of the twinning process, taking into account that the melting point of the Zinc is much more lower than the other materials investigated by previous studies.

The micropillar compression experiments were performed using a custom-made nanoindenter equipped with a flat punch doped diamond tip with a diameter of $13\mathrm{\:\mu m}$. The tip was attached to a spring with a spring constant of $1.7$~mN/$\mu$m that was used to determine the acting force. During the compression no load or strain feedback was implemented rather the platen that held the other end of the spring was moved with a constant $10\mathrm{\:nm/s}$ indentation speed. The reason for not using stress- or strain-control was two-fold: i) As known from earlier investigations the intermittent dynamics of dislocations takes place on a time-scale of $\mu$s-ms, and the feedback loop can only affect dynamics on a larger timescale (since the acquisition rate of force and displacement measurements was 200 Hz) and ii) Stress- and strain-control would require the instantaneous quick movement of the platen that may lead to an undesired elastic response of the coupled sample-device system that might even trigger artefact AE signals and/or strain bursts.

\subsection{Detecting acoustic emission}
\label{sec:method_ae}

The device was also equipped with a piezoelectric AE transducer that allowed concurrent streaming of the AE signal during experiments carried out in-situ inside the vacuum-chamber of the SEM. The sample hosting the micropillars was clipped to the transducer surface where a layer of vacuum grease was applied to ensure effective coupling~\cite{Hegyi.2017,Kalacska.2020, ispanovity2022dislocation}. The piezoelectric transducer used for AE detection was a Physical Acoustics Corporation Micro30S wide-band (100-1000 kHz) AE sensor. The recorded signal was amplified using the Vallen AEP5 pre-amplifier set to $40~\mathrm{dB_{AE}}$. Data acquisition was performed using the computer-controlled Vallen AMSY-6 system and was carried out in continuous data streaming mode at a sampling rate of $2.5~\mathrm{MHz}$.

To individualize the AE events from the measured continuous $V(t)$ signal the threshold voltage was set to $V_\mathrm{th} = 0.018$~mV, this value being slightly above the background noise. The hit definition time (HDT), i.e.,~the minimum period between two subsequent AE events was $20~\mathrm{\mu s}$.

\begin{figure}[t!]
	\includegraphics[width=0.48\textwidth]{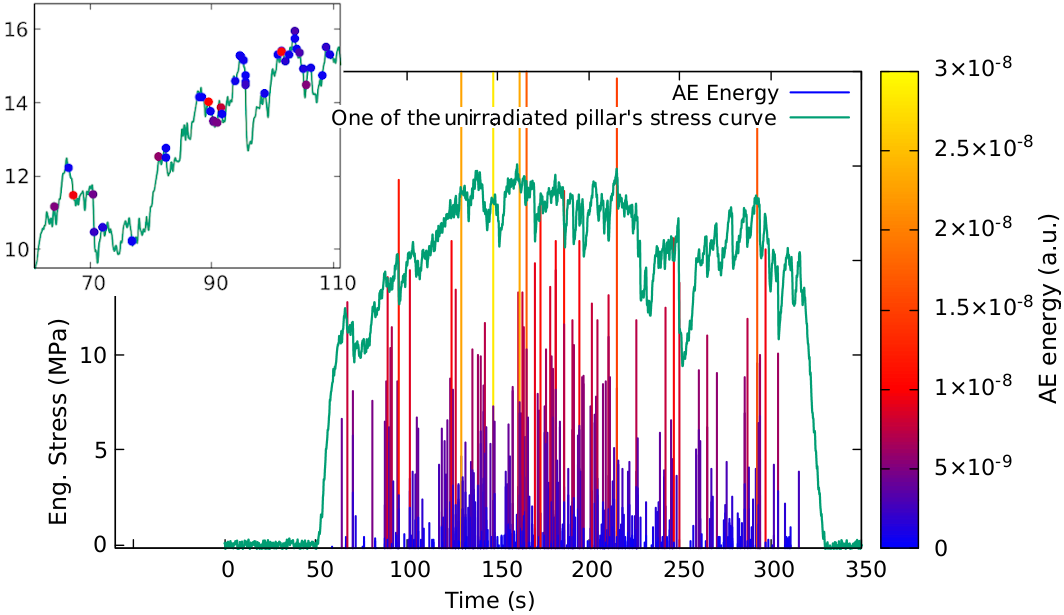}
	\caption{The stress vs.~time curve (green line) obtained during compression of the unirradiated pillar of Fig.~\ref{fig:ebsd}. The impulse signals represent the detected AE events with their heights as well as their colours referring to their energies as defined by the right axis and the colourbar, respectively. To emphasize the match between the AE events and the stress drops the inset shows a zoomed part of the curve with the AE events indicated as dots with the same colour coding as on the main panel.}
	\label{fig:stresstime}
\end{figure}

The stress vs.~time recorded during the compression of an unirradiated micropillar is shown in Fig.~\ref{fig:ebsd} together with the detected AE events is shown in Fig.~\ref{fig:stresstime}. The inset indicates that a perfect match between the AE events and the detected stress drops was obtained.

%\picomment{Dávid: ide le kellene írni az AE kísérlet részleteit, milyen a kütyünk, hány db, ilyesmi...}

\section{Experimental results}

%\textcolor{cyan}{Ez mondjuk részletezi a szimulációs részét
%Does irradiation enhance or inhibit strain bursts at the submicron scale? Yinan Cui Giacomo Po, Nasr Ghoniem http://dx.doi.org/10.1016 /j.actamat.2017.04.055}

\subsection{Micromechanical properties}

%\picomment{Az első bekezdést egyelőre benthagytam, de nem tudom mi legyen majd vele.} Although some studies~\cite{WEAVER2017368, REICHARDT2017323, REICHARDT2015147} 
%\textcolor{cyan}{Érdemes még bővíteni az ilyen cikk felsorolást? Ez mondjuk self-ion Fe besugárzott "ronda" pillárokon látta, hogy az unirradiated drop-jai nagyobbak (de nem tette szóvá)
%Compression of self-ion implanted iron micropillars
%E.M. Grieveson, D.E.J. Armstrong
%, S. Xu, S.G. Roberts}
As it was mentioned in the introduction, there is no consensus in the literature whether irradiation enhance or inhibit strain bursts \cite{cui2017does}. Obviously, a general answer cannot be given to this question, as the dynamics of dislocations that is responsible for the strain bursts may depend on the crystal structure, its orientation, type and distribution of the irradiation-induced faults, temperature, etc. To simplify the problem, we consider a single slip scenario where forest dislocations are practically absent and a moderate level of irradiation that is expected to introduce mostly dislocation loops. In addition, to exclude the external influences of the experimental setup on the strain bursts we do not apply any stress or strain feedback (since the former could lead to the coalescence of strain bursts and the latter may lead to unnecessary unloading of the sample after larger events), but rather a constant platen velocity is imposed as described in Sec.~\ref{sec:micromechanical_experiments}.
%In this study we were trying to exclude the external influences of the experimental setup as the deformation mode by using nanoindentation device without any stress or strain feedback, and the complex and complicated dislocation interaction by applying micropillars oriented to single slips. %\picomment{Itt akkor arról kéne dűlőre jutni, hogy korábban azt látták hogy nőnek vagy csökkennek a dropok? Nyilván ez nagyban függ a deformációs módtól, pl. feszültségvezérléstől simán nőhetnek alakítási puhulás esetén.}

\begin{figure}[t!]
	\includegraphics[width=0.48\textwidth]{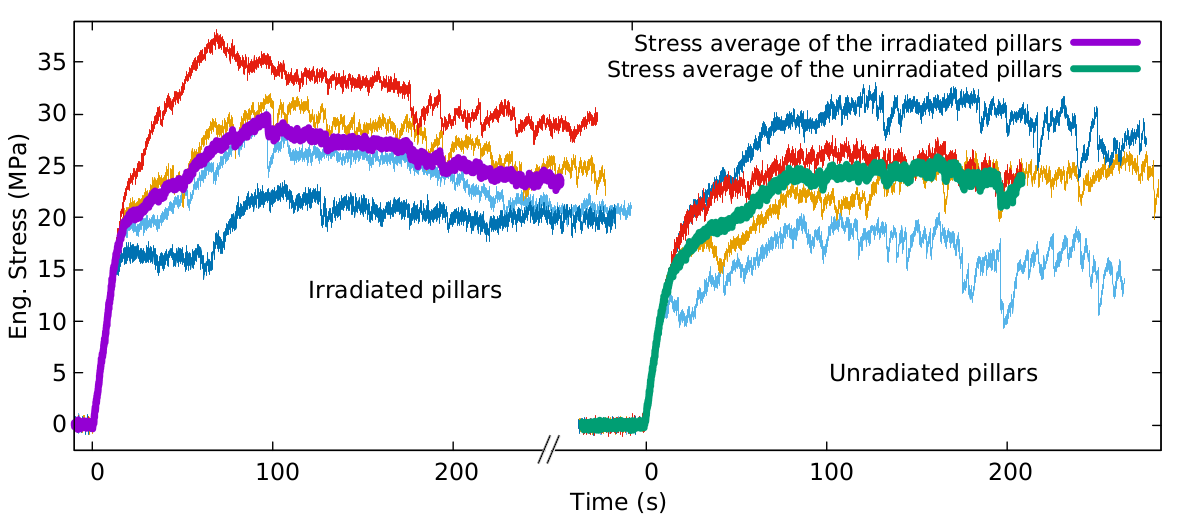}
	\caption{Experimental stress vs.~time curves grouped in two sets (irradiated and unirradiated). The calculated average curves for the two sets are plotted with thick purple and green lines.}
	\label{fig:strtime}
\end{figure}

\begin{figure}[t!]
\begin{centering}
	\includegraphics[width=0.4\textwidth]{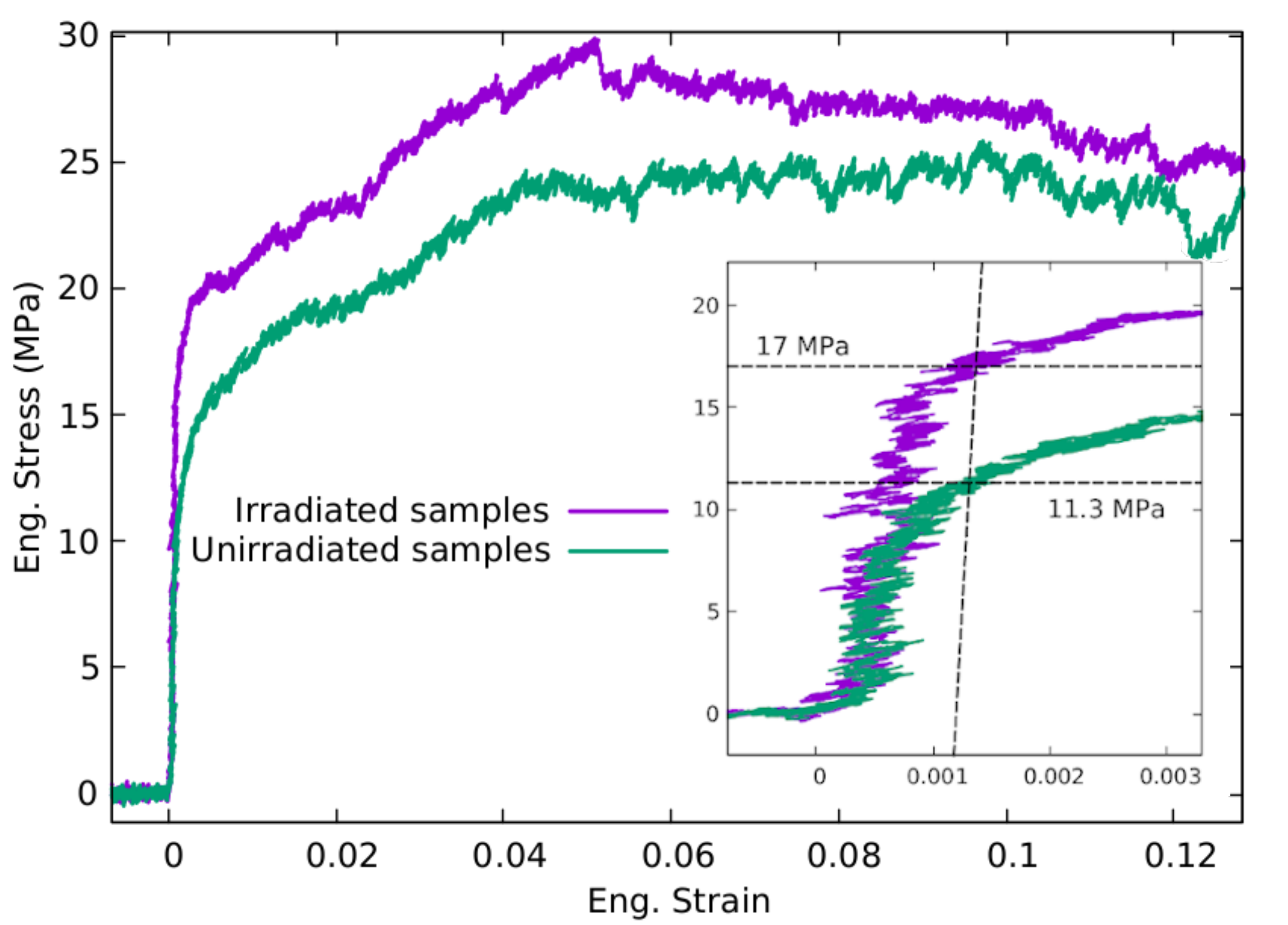}
	\caption{Average stress-strain curves of the unirradiated and irradiated pillars. The inset shows how the average yield stress was determined for the two cases. Notice the hardening caused by the irradiation.}
	\label{fig:strstr}
\end{centering}
\end{figure}

The stress vs.~time curves measured for the unirradiated and irradiated pillars are shown in Fig.~\ref{fig:strtime}. It is evident that, as expected, large stress drops dominate the deformation process already from the end of the purely elastic regime. The stochastic behaviour is manifest in the fact that curves in both sets differ from each other due to the different (though statistically equivalent) initial dislocation structures. By averaging the individual measurements one can arrive at a curve statistically representative for the micropillars of the given size~\cite{ispanovity2013average}. These average curves are also shown in Fig.~\ref{fig:strtime} as well as the corresponding stress-strain curves in Fig.~\ref{fig:strstr}. As one can notice, hardening caused by irradiation can be clearly observed. According to the inset the increase in the yield stress is around 50\%: $\Delta \sigma_\mathrm{y} \approx 5.7$~MPa. Since according to the TEM studies hardening is due to small dislocation loops, the dispersed barrier-hardening model of Eq.~(\ref{eqn:hardening}) is expected to apply. Indeed, if material characteristics of Zn ($b=0.266\mathrm{\:nm}$ and $G=3.5\mathrm{\:GPa}$), the measured parameters of the loops ($N = 2.4\mathrm\:\times 10^{-7} \mathrm{nm}^{-3}$ and $d= 105\mathrm{\:nm}$) and the calculated Schmid factor of $m=0.48\pm 0.02$ is used then $\alpha=0.61\pm 0.15$ is obtained for the strength factor. This value of $\alpha$ suggests that the loops act as strong barriers for dislocation motion~\cite{lucas1993evolution}. In general, dislocation loops are considered week barriers, since mobile dislocations are usually able to bypass them via cross-slip, however, this mechanism is suppressed at room temperature in the highly anisotropic HCP structure of Zn that can explain the stronger pinning effect of the loops.

\begin{figure}[t!]
    \includegraphics[width=0.4\textwidth]{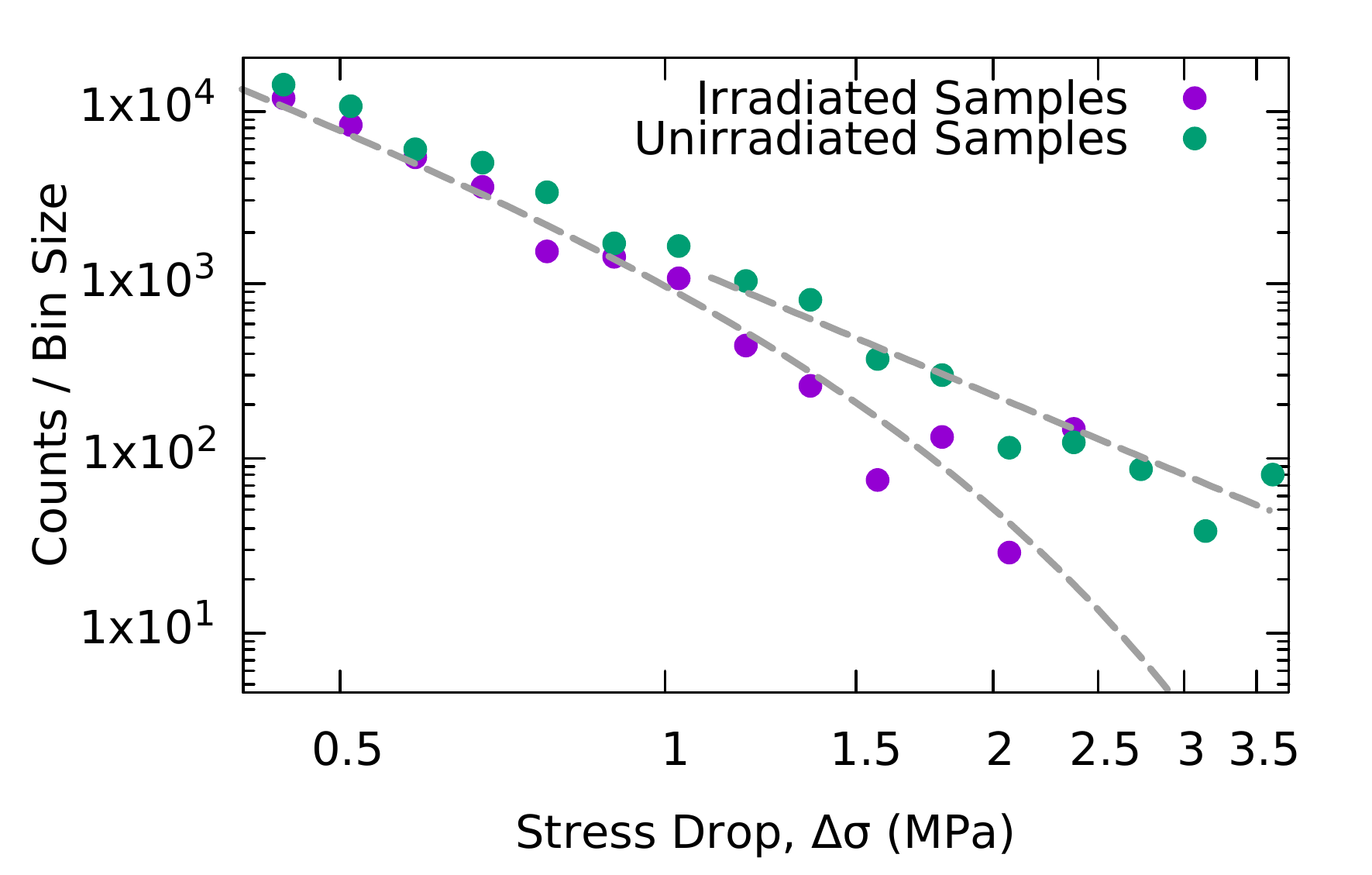}
    \caption{Distribution of the stress drop sizes before and after irradiation. Instead of the probability density function the counts per bin size is plotted, which differ only in a multiplicative factor. The reason is, that this way the values represent the total number of stress drops with a given magnitude (since equal number of micropillars were compressed in the two cases). It is seen that the number of small ($\lesssim 1$MPa) stress drops remains the same but a strong decrease is seen in the number of large drops due to irradiation (as can be also seen visually in Fig.~\ref{fig:stresstime}). The data were fitted with $P(\Delta\sigma) \propto \Delta\sigma^{-\tau}\cdot e^{-(\Delta\sigma/\sigma_{0})}$ (dashed lines), where $\tau=2.4$, $\sigma_{0}=1.2$ (irradiated case) and $\sigma_{0}=\infty$ (unirradiated case).
    \label{fig:stress_drop}
    }
\end{figure}

According to Fig.~\ref{fig:strtime} the stress drops characteristic of micron scale plasticity prevail after irradiation. However, visual inspection suggests that the drop sizes decrease after ion implantation. Indeed, the distributions of the drop sizes plotted in Fig.~\ref{fig:stress_drop} confirm that the number of stress drops below $\sim$1 MPa does not change, whereas larger drops appear in a significantly smaller amount after ion implantation. Although the limited amount of data introduces a large scatter in the distributions, it seems plausible that the dispersed dislocation loops created by the irradiation introduce an exponential cut-off in the otherwise scale-free distribution. This result is in line with the results of Zhang et al.,~where the effects of both specimen size and different obstacles on the strain burst size distribution were investigated in Al alloys \cite{zhang2017taming}. It was found that increase in obstacle density reduces the fluctuations, a phenomenon expressed as `dirtier is milder'.
%In the current measurements the noise due to the limited number of the total detected stress drops prevents a qualitative analysis of the cut-off in Fig.~\ref{fig:stress_drop}, Nevertheless, it is evident that the typical stress drop sizes decrease that we attribute to the appearance of a length-scale due to the dispersed dislocation loops created by the irradiation.

Finally, it is noted that in the unirradiated case strain hardening is neither seen on the average stress-time curve nor on the individual ones in Fig.~\ref{fig:strtime}. This is consistent with previous investigations on the same material and it is attributed to the absence of forest dislocations and dislocation reactions due to the single slip type of deformation~\cite{ispanovity2022dislocation}. After irradiation, however, a slight strain softening is observed. The effect is apparent not only on the averaged curve but also on all individual curves of the micropillars in Fig.~\ref{fig:strtime}. The possible reason for this behaviour can be that reactions of mobile dislocations with the static loops may lead to slip bands with a reduced number of barriers, a phenomenon called dislocation channelling \cite{greenfield1961effect, victoria2000microstructure, farrell2004deformation}, leading to strain localisation and softening. In the next section, therefore, we analyse whether strain localisation can be indeed observed during the compression of irradiated micropillars.

%We note, however, that no clear signs of strain localisation could be detected on the post mortem SEM images of compressed irradiated pillars (e.g., Fig.~\ref{fig:ebsd}).

%\picomment{Is there some evidence of channelling on the SEM images of irradiated pillars?}

\subsection{Strain localisation}
\label{sec:localisation}

As it was discussed above, the micromechanical features suggest that strain localisation causing strain softening may be present in irradiated samples. In this section a quantitative analysis of this phenomenon is provided based on the SEM images taken during the microcompression experiments.

\begin{figure}[t!]
	\includegraphics[width=0.48\textwidth]{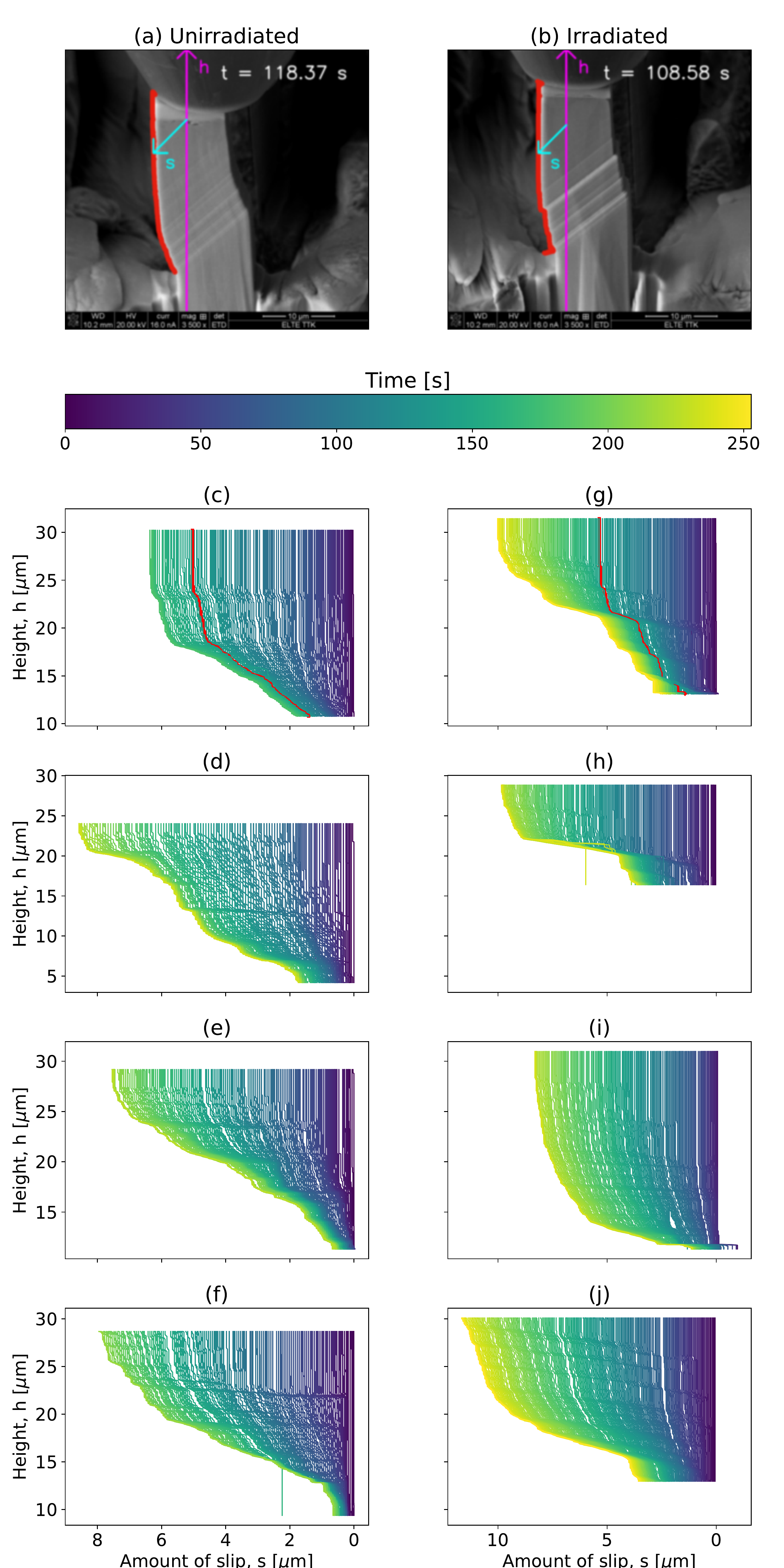}
	\caption{Evolution of micropillar shapes during compression experiments. Left column: unirradiated pillars, right column: irradiated pillars. The SEM images [panels a) and b)] provide examples how the micropillar edges are found with image processing and the definition of the $h$ and $s$ axes. Panels c)-j) show the evolution of the edge geometries for all compressed pillars for different times represented by the colour scale. The two red curves in panels c) and g) correspond to the edges found in the SEM images in panels a) and b), respectively. % \vspace*{-2cm}
%	Snapshots of micropillars during microcompression for two representative experiments. Upper row: unirradiated pillar. Bottom row: irradiated pillar.
%	The instants when the images were taken are labeled with roman numbers on the stress-time curves shown on the right-hand-side. Note, that i) a distinct shear band forms in the case of the irradiated pillar in the strain softening regime and ii) no such strain localisation occurs in the unirradiated case. In order to analyse the evolution of the pillar shape, the left edge of the pillar was detected using image processing algorithms, and are highlighted with different colours. 
	\label{fig:localisation}}
\end{figure}

Figure \ref{fig:localisation} is concerned with the evolution of the specimen shape during compression. For the purpose of characterising the shapes, SEM images recorded with a backscattered electron detector at a rate of 1 FPS during the compression were used. The position of the left edge of the pillar was detected using image processing since it exhibits a large contrast with respect to the background. To this end we utilised the open source OpenCV library.

As an initial step we searched for edges with the Canny method. The two threshold levels for lower and upper brightness were selected to be 150 and 220, respectively, on the 0--255 scale. After that each point on each image which were found by the Canny algorithm were identified as potentially belonging to the pillar edge. In order to eliminate the contour of other background features, a recursive procedure based on minimizing neighbor-to-neighbor distances was applied.
%
%After that we located a point on each image which is part of the edge and by selecting points which were also found by the Canny algorithm as an edge and by applying the same rules to them recursively while there are no more points we located the edge. These point were projected to a reference line $h$.
%
To quantify the shape a tilted coordinate system was employed to account for the fact that slip only occurred on the basal plane having an inclination of $\approx45^\circ$ with the vertical loading axis. The points of the edge were, therefore, projected to the vertical axis $h$ with 45$^\circ$ angle and the value $s(h)$ expresses the amount of displacement that took place at a height of $h$.
%(see top panels in Fig.~\ref{fig:localisation} for the definitions of the coordinate axes $h$ and $s$).
Panels a) and b) of Fig.~\ref{fig:localisation} show examples of a pristine and an irradiated pillar, how the edge of the pillar was found and how the shapes $s(h)$ are obtained. The time evolution of $s(h)$ during all of the microcompression experiments is visualized below in Fig.~\ref{fig:localisation}c-j. The specific shapes corresponding to images of panels a) and b) are highlighted with red colour.

In the unirradiated case (Fig.~\ref{fig:localisation}, left column) the slip surface is inhomogeneous, as expected, yet, the slip bands (seen as jumps on the $s(h)$ curves) are still distributed quite evenly along the height of the micropillar. In contrast, in the irradiated pillars slip bands appear homogeneously only during the initial stages of the deformation after which deformation localises mostly to one or few slip bands with approx.~100 nm thickness. These observations are also corroborated by the Supplementary Videos 1-8 showing the course of deformation for all individual experiments.

\begin{figure}[t!]
	\includegraphics[width=0.48\textwidth]{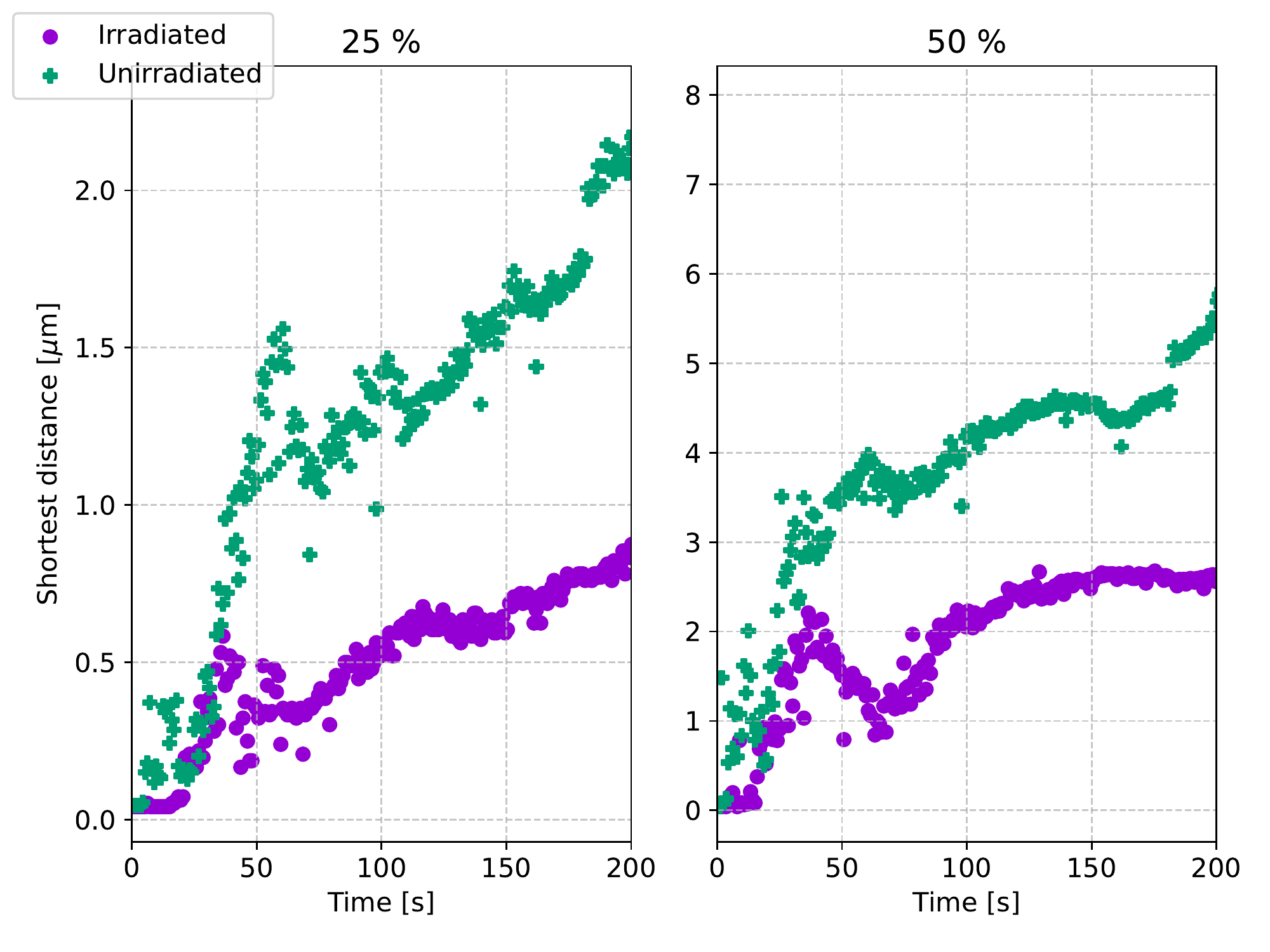}
	\caption{Localisation width $w_\text{min}(p)$ [see Eq.~(\ref{eqn:w_min})] for $p$ values chosen as 25\% (left panel) and 50\% (right panel) of the total plastic slip of the micropillar. The values were averaged for both the pristine and irradiated pillars.
	\label{fig:localisation_w}}
\end{figure}

In order to quantify the visually observed localization in the irradiated case, the minimum width of a shear band was determined using the shapes $s(h)$ in which a pre-defined value of plastic slip $p$ is realized, formally,
\begin{equation}
    w_\text{min}(p) = \min_{h} \left[ s^{-1}(s(h)+p)-h \right].
    \label{eqn:w_min}
\end{equation}
Here we exploited the fact that $s(h)$ is a monotonically increasing function due to the single slip type of deformation. According to this definition, the smaller the value of $w_\text{min}(p)$ is, the more is deformation localised into a single slip band. Since the total plastic slip $S=s(L)-s(0)$ increases monotonically in time, in order to study the localization $p$ is chosen to be proportional with $S$. Figure \ref{fig:localisation_w} plots the time evolution of $w_\text{min}(p)$ for pristine and irradiated pillars where $p$ is either $S/4$ or $S/2$ (denoted as 25\% and 50\%, respectively). As seen the widths $w_\text{min}(p)$ initially increase linearly, showing that deformation is less and less local, that is, new slip bands appear at different positions. This increase slows down considerably as strain increases indicating some localization in both cases. It is evident, however, that the localization is significantly stronger for irradiated pillars as the width of the slip band containing either 25\% or 50\% of the total strain is approximately halved, so deformation is much more concentrated into a narrow slip band in this case. The localisation is clearly related to the strain softening discussed in the previous section and its physical origin is expected to be dislocation channelling, that it, due to the reactions between irradiation loops and mobile dislocations the obstacle density locally decreases, thus reducing the strength in the plastically deformed regions.

%\begin{figure}[th!]
%%	\includegraphics[width=0.48\textwidth]{figu%res/wtpdf.pdf}
%	\caption{}
%\end{figure}

%To quantify localization, first, the shape of the pillar was detected. For this purpose the SEM images recorded with a backscattered electron detector (BSD) at a rate of 1 FPS were used. With the specific settings used the left edge of the micropillar was much lighter than the background which enabled ... \picomment{We need input from Gábor}.

%For an unambigous determination 

\subsection{Acoustic emission}

\begin{figure}[t!]
	\includegraphics[width=0.48\textwidth]{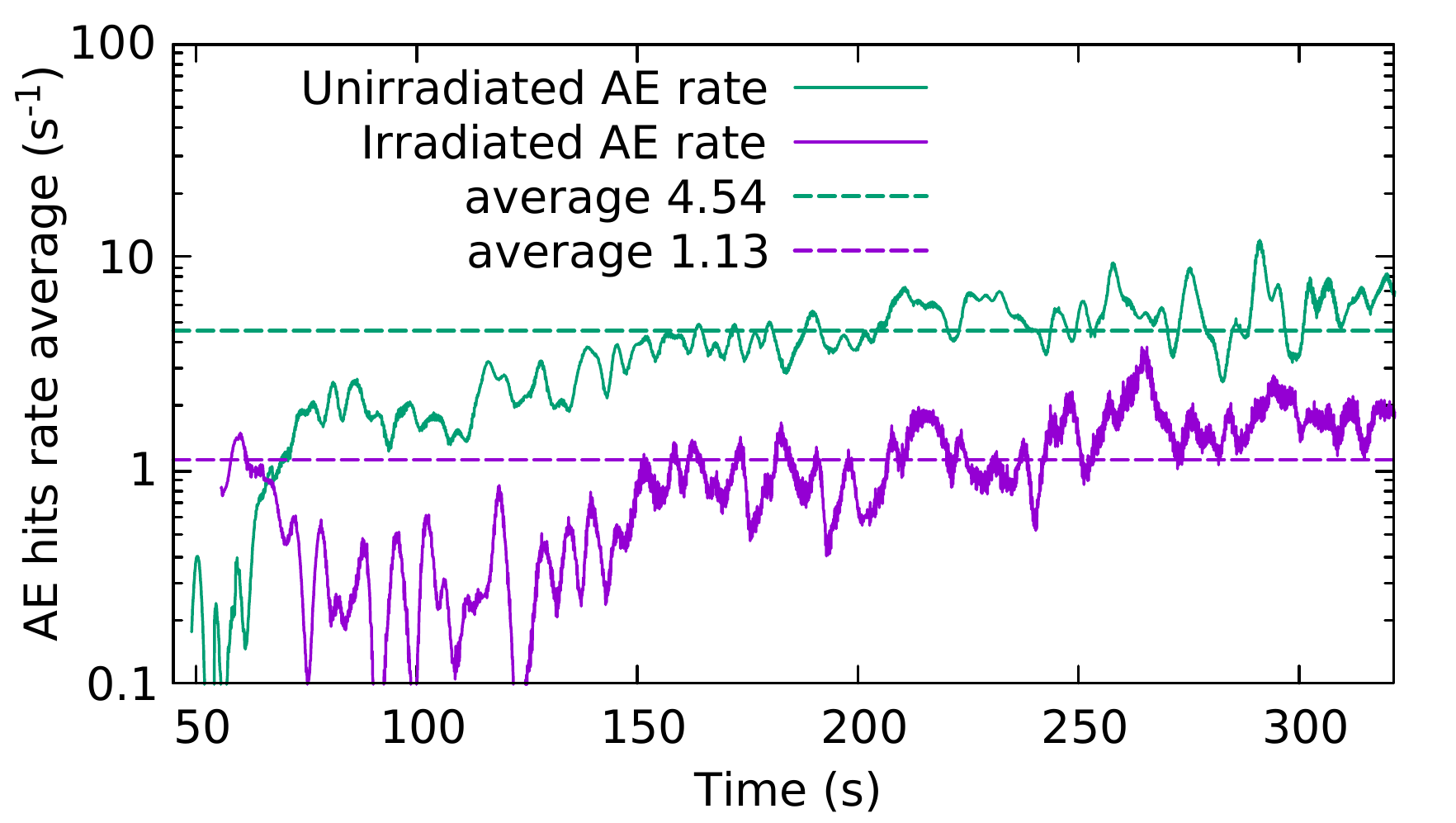}
	\caption{The rate of the detected AE events averaged over the individual measurements on irradiated and unirradiated micropillars. The rate was obtained by convolving the individual hits with a Gaussian with FWAHM\,=\,1\,s.
	\label{fig:rateave}}
\end{figure}

As already mentioned in Sec.~\ref{sec:method_ae}, a large number (typically several hundreds) of individual AE events could be detected during the compression of each pristine and irradiated micropillar. In both cases the events were correlated with the stress drops that correspond to the stochastic strain burst events (see Fig.~\ref{fig:stresstime}). The rate of the events, however, is decreased with a factor of $\sim$4 due to irradiation as seen in Fig.~\ref{fig:rateave}. The difference in the rate is even larger during the first approx.~150\,s of the compression, but otherwise no strong time dependence is seen, the rates are nearly constant throughout the deformation. This observation can be again attributed to the possible onset of dislocation channelling, that it, the increase of the defect-scarce regions gives rise to more extensive correlated dislocation movements.

\begin{figure}[t!]
	\includegraphics[width=0.48\textwidth]{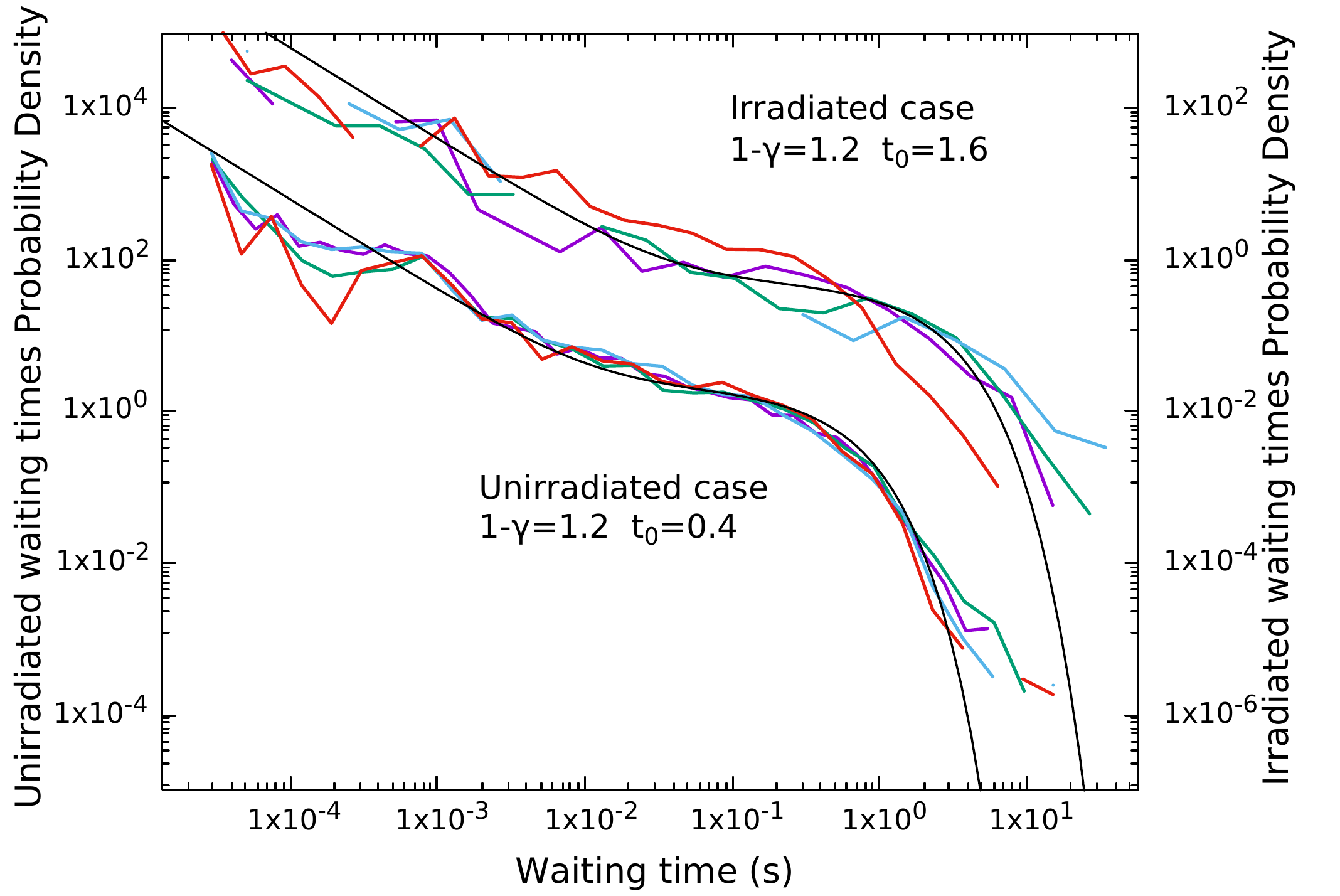}
	\caption{Probability distribution of waiting times between consecutive AE events. Data from irradiated and unirradiated cases are shifted along the $y$ axis with two orders of magnitude to make differences visible. The fitted curves correspond to Eq.~(\ref{eqn:waiting_time}) with parameters indicated in the figure.
	\label{fig:waiting_time}}
\end{figure}

\begin{figure}[t!]
	\includegraphics[width=0.4\textwidth]{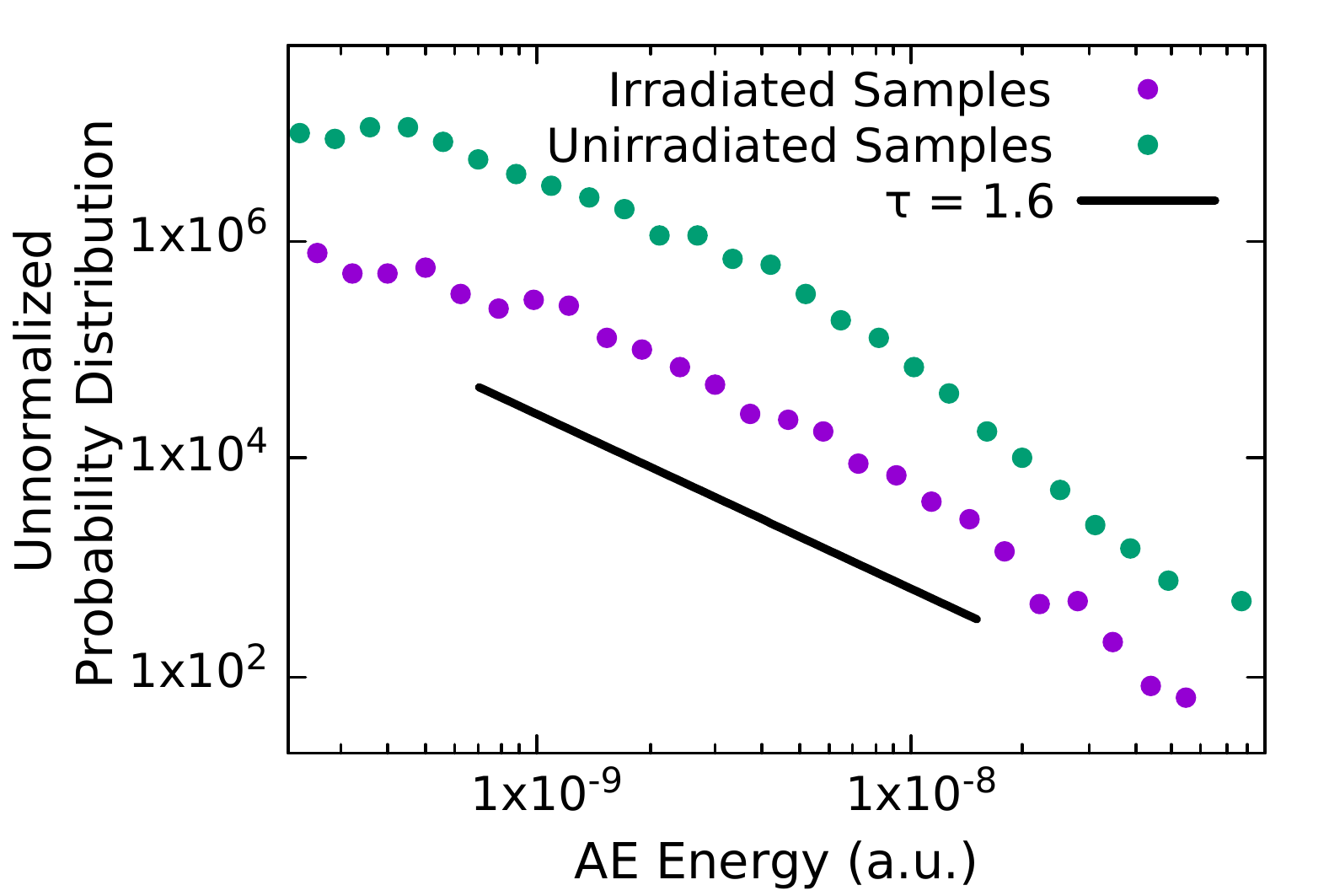}
	\caption{Probability distributions of the emitted AE energy from all irradiated and unirradiated pillars. The thick black line denotes the power-law slope of $\tau=1.6$.
	\label{fig:AEpdf}}
\end{figure}

In order to obtain deeper insights regarding the dislocation activity, we turn to the analysis of the distribution of waiting times $t_\mathrm{w}$ between subsequent AE events plotted in Fig.~\ref{fig:waiting_time}. The shape of the obtained distributions $P(t_\mathrm{w})$ is equivalent to that found earlier for pristine Zn pillars \cite{ispanovity2022dislocation}
\begin{equation}
    P(t_\mathrm{w}) = [A t_\mathrm{w}^{-(1-\gamma)} + B]\exp(-t_\mathrm{w}/t_0).
    \label{eqn:waiting_time}
\end{equation}
This distribution consists of two separate regimes: at short waiting times it is characterized by a power-law decay with exponent $(1-\gamma)$ whereas at longer waiting times an exponential cut-off is seen at the characteristic timescale of $t_0$. The detailed analysis of \cite{ispanovity2022dislocation} revealed that short waiting times account for correlated AE clusters which originate from the same stress drop and can be thought of as main shock-aftershock sequences of earthquakes. On the other hand, longer waiting times are characteristic of signals belonging to different stress drops, they are uncorrelated and, therefore, can be described as a Poisson process. According to Fig.~\ref{fig:waiting_time} the shape of the distribution is unchanged after irradiation, so the correlated clusters of AE events prevail, and no significant change in the dynamical behaviour is observed. However, the decrease in the AE activity mentioned above is manifest from the increase of the time-scale $t_0$, where the same 4$\times$ ratio is observed as in the drop of average AE event rate. This suggests, that the dynamics of dislocations during an avalanche as can be detected by AE does not change, however, the number of these complex collective dislocation events drop with a factor of approx.~4. These findings are corroborated by the distribution of the released energies of individual AE events plotted in Fig.~\ref{fig:AEpdf} since no visible change is seen in the shape before and after irradiation. In conclusion, the acoustic signal in both cases consists of correlated event clusters with identical parameters. However, irradiation reduces the number of these clusters (with a factor of $\sim 4$), especially at the beginning of deformation, where the decrease is even stronger. 
%\picomment{itt esetleg érdemes lenne megnézni, hogy milyen különbségek vannak az első 150 s és a második 150 s között...}

%\ducomment{Ez nem tök jól magyarázza azt is, hogy a besugárzott esetben annak ellenére hogy kevesebb az AE, az AE "cutoffja" mégsincsen annyival lejebb?}.

%\textcolor{cyan}{In situ TEM study on elastic interaction between a prismatic loop and a gliding dislocation
%S.Liua
%Sok dl-loop kölcsönhatás} \picomment{Ezt pontosan hova tennéd be?}

%Though, it seems that the distribution of the AE independent from the irradiation, the number of the detected events are so sensitive for that. To examine this phenomena it was calculated the AE hits rate. The rate vs. time could be given by convoluted the signals - replaced them velue to 1 - with gaussian function (FWAHM=1). This method represent the hits rate average perfectly and adequately characterized it's change in time.

\section{Discussion}

As known from earlier measurements on ice and Zn single crystals plastic fluctuations in materials with HCP structure tend to be wild in the sense that plastic instabilities due to dislocation avalanches can be observed at any sample size even up to bulk scales and the plastic events are accompanied by strong acoustic emission signals \cite{becker1932, miguel2001intermittent, weiss2007evidence, ispanovity2022dislocation}. The shear bands associated with the plastic events are usually distributed evenly along the length of the sample, indicating that the deformation in a specific shear band causes local hardening and the subsequent shutdown of plastic activity in that band. In this paper our main aim was to investigate how defects induced by irradiation change this picture. A further goal was to measure how the presence of obstacles with short-range forces change the collective dynamics of dislocations in therms of the AE signals.

To prepare microsamples with a homogeneous damage structure a multi-energy p$^+$ irradiation scheme was applied that led to a radiation damage of approx.~0.05 dpa. Microstructural analysis performed using TEM and XRD revealed that mainly dislocation loops on the basal plane were formed with diameters up to approx.~100 nm. These act as obstacles against basal glide of mobile dislocations and, therefore, lead to hardening the amount of which was found to be quantitatively consistent with the dispersed barrier-hardening model, as expected.

One of the most prominent features caused by the irradiation was the localization of deformation into one or few slip bands, a phenomenon that was observed earlier for other types of materials, too. Here, based on the stress-time curves and the micropillar shapes, we provided an in-depth quantitative analysis of the how localization takes place, and concluded that if threshold strain in the slip band is reached then strain softening sets in and leads to the localization. This is consistent with the picture, that the interaction of mobile dislocations and the irradiation loops reduce the defect density in the slip band leading to local softening (contrary to the situation in pristine pillars, explained above).

As of the dynamics of the dislocations during the plastic events, an interesting picture has emerged. According to the stress drop statistics extracted from the stress-strain curves the irradiation leads to the appearance of a cut-off in the distribution, that is, a characteristic maximum size of strain bursts is introduced. This effect can also be observed visually on the force-time curves of Fig.~\ref{fig:stresstime}. This seems to be a straightforward consequence of the presence of dislocation loops due to irradiation since static obstacles with short-range stress fields are known to inhibit strain bursts \cite{zhang2017taming}. In such a case, dislocations get pinned at the obstacles and the dynamics which is governed by long-range elastic interactions of dislocations in pure systems is gradually replaced by the dominance of short-range dislocation-obstacle interactions. This effect was thoroughly studied also with 2D and 3D discrete dislocation dynamics simulations \cite{ovaska2015quenched, salmenjoki2020plastic, berta2022dynamic}. In the present case, however, the situation is different as can be concluded from AE measurements. Somewhat surprisingly, no difference was observed between the energy distributions and the main features of the waiting time distributions in the unirradiated and irradiated cases. In particular, the waiting time distributions report about identical scale-free correlations between subsequent AE events belonging to the same cluster that were previously identified as main shocks and aftershocks similarly to earthquake dynamics \cite{ispanovity2022dislocation}. In addition, not cut-off was found to appear in the distribution of AE energies. Since AE signals are considered as fingerprints of the collective dynamics of dislocations, we conclude that large intermittent dislocation avalanches characteristic of pure metals and single slip deformation prevail after irradiation. However, a significant difference was seen in the rate of the AE events, namely, irradiation lead to a decrease of the number of events with a factor of approx.~4. One may thus conclude that a mechanism other than large collective dislocation avalanches may be also active that can be responsible for a large portion of the plastic strain. We speculate that this could be the 'small' dislocation avalanches due to dislocation-obstacle interactions, that is, unpinning and pinning of single dislocation lines. During these events the released elastic energy is much smaller than for collective avalanches that could explain why these events are not detected by AE measurements.

So, the AE measurements do not provide direct explanation for the decrease in the strain burst sizes due to irradiation, since the energies and waiting times remain practically unchanged. As it was discussed previously as an effect of irradiation a significant strain softening was observed that caused dislocation channeling and lead to the localization of plastic strain to narrow shear bands. Thus, in the irradiated samples the extension of the dislocation avalanches were not only constrained by the width of the micropillar but also by the width of the shear band. Since internal length-scales are known to limit the extension of dislocation avalanches, this new length-scale may be responsible for the cut-off seen in the stress drop distribution. It is stressed, however, that due to strain softening the distribution is expected to be very sensitive to the applied deformation mode. In particular, with stress control a single strain burst leading to failure is expected after reaching the peak stress. So, according to our experiments whether irradiation promotes or inhibits strain bursts is dependent on the mode of loading.

In conclusion, three individual measurement methods (SEM imaging, stress-displacement measurements and detection of AE signals) were performed simultaneously during the compression of pristine and irradiated Zn micropillars. The experiments revealed a complex dynamic behaviour of dislocations that was affected in many aspects by the formation of irradiation-induced defects, mostly dislocation loops. Firstly, the yield stress increased due to the introduced obstacles. Secondly, the dislocation-obstacle interactions upon plastic strain lead to the formation of thin dislocation channels that carried most of deformation at the second stage of compression. The formation of these channels were accompanied by strain softening, an effect attributed to the removal of obstacles from the channels. Thirdly, strain burst sizes were found to decrease after irradiation that was speculated to be caused by the finite thickness of the dislocation channels that posed a constraint to the spatial extension of dislocation avalanches. Finally, the analysis of AE events did not report any fundamental difference in the fine dynamic structure of dislocation avalanches, however, their total number was significantly decreased, likely because some plastic activity also took place outside dislocation channels.

\section*{Acknowledgments}
%DU was Supported by the ÚNKP-21-4 New National Excellence Program of the Ministry for Innovation and Technology from the source of the National Research, Development and Innovation Fund.

Financial support from the National Research, Development and Innovation Fund of Hungary is acknowledged under the young researchers’ excellence programme NKFIH-FK-138975 (D.U., G.P., S.L., I.G.~and P.D.I) and under the ÚNKP-21-4 (D.U.) and ÚNKP-21-3 (G.P.) New National Excellence Programme of the Ministry for Innovation and Technology.
Authors thank the support of VEKOP-2.3.3-15-2016-00002 of the European Structural and Investment Funds (Zs.F.). This paper was also supported by the János Bolyai Research Scholarship (Zs.F.) of the Hungarian Academy of Sciences. The authors thank L.~Illés for FIB preparation of the TEM lamella. Further financial and technical support was provided by Metlok Engineering Kft,~the exclusive distributor of the vacuum feedthroughs of Allectra GmbH.

\footnotesize{
\bibliography{rsc} %your .bib file
\bibliographystyle{rsc} %the RSC's .bst file
}
\end{document}